\documentclass[pdftex
]{eptcs}
\usepackage[T1]{fontenc}
\usepackage{enumerate,picins}
\providecommand{\event}{SOS 2010}
\title{Structural Decomposition of Reactions \\
  of Graph-Like Objects}
\author{Tobias Heindel \institute{  Abteilung
    f\"ur Informatik und angewandte 
    Kognitionswissenschaft
    \\ 
    Universit\"at Duisburg-Essen, 
    Duisburg, 
  Germany}}
\def\titlerunning{Structural Decomposition of Reactions}
\def\authorrunning{T. Heindel}
\usepackage{calc}
\usepackage{breakurl}
\usepackage{graphics,microtype,xspace}
\usepackage{stmaryrd,amssymb,marvosym}
\usepackage{url}
\usepackage{ifthen}
\usepackage{subfig}

\newenvironment{mycolor}[1]{\color{#1}}{}

\newcommand{\todo}[1]{}
\usepackage[mathscr]{euscript}
\newcommand{\id}{\mathrm{id}}
\newcommand{\op}{\mathrm{op}}

\newcommand{\Sub}[1][]{{\mathrm{Sub}_{\mathscr{#1}}}}

\newcommand{\mnode}[3][ ]{%
  \ifthenelse{\equal{#1}{ }}%
  {\node[inner sep=2pt] (#2) at (#3) {$#2$};}%
  {\node[inner sep=2pt] (#1) at (#3) {$#2$};}
}

\newcommand{\POCS}[4][.27]{%
  \begin{scope}[scale=#1]
  \path[draw=none] (#2) -- ++(#3) coordinate (myXPOCone);    
  \path[draw=none] (#2) -- ++(#4) coordinate (myXPOCother);
  \draw[shorten <=1.5pt] (myXPOCone) -- ++(#4);
  \draw[shorten <=1.5pt] (myXPOCother) -- ++(#3);
  \end{scope}
}
\newcommand{\POCSd}[4][.2]{%
  \begin{scope}[scale=#1]
  \path[draw=none] (#2) -- ++(#3) coordinate (myXPOCone);    
  \path[draw=none] (#2) -- ++(#4) coordinate (myXPOCother);
  \draw[shorten <=1.5pt,white,ultra thick] (myXPOCone) -- ++(#4);
  \draw[shorten <=1.5pt,white,ultra thick] (myXPOCother) -- ++(#3);

  \draw[shorten <=1.5pt] (myXPOCone) -- ++(#4);
  \draw[shorten <=1.5pt] (myXPOCother) -- ++(#3);
  \end{scope}
}

\newcommand{\POC}[3][.4]{%
  \path[draw=none] (#2) -- ++(#3:#1*1.4142136) coordinate (myXPOC);
  \path[draw=none] (#2) -- ++(#3+45:#1) coordinate (myXPOCone);
  \path[draw=none] (#2) -- ++(#3-45:#1) coordinate (myXPOCother);
  \draw[shorten >=1.5pt] (myXPOC) -- (myXPOCone);
  \draw[shorten >=1.5pt] (myXPOC) -- (myXPOCother);
}

\newcommand{\ardrawl}[5][--]{%
  \begin{scope}[every node/.style={inner sep=2pt}]    
    \draw[->] (#2) #1 node[auto,#4] {$\scriptstyle #5$} (#3) ; 
  \end{scope}
}
\newcommand{\ardrawld}[5][--]{%
  \begin{scope}[every node/.style={inner sep=2pt}]    
    \draw[ultra thick,white] (#2) #1 (#3);
    \draw[->] (#2) #1 node[auto,#4] {$\scriptstyle #5$} (#3) ; 
  \end{scope}
}
\newcommand{\ardraw}[3][--]{%
  \ardrawl[#1]{#2}{#3}{}{}%
}

\newcommand{\ardrawLd}[5][--]{%
  \draw[ultra thick,white] (#2) #1 (#3);
  \ardrawl[#1]{#2}{#3}{#4}{#5}
}

\usepackage{amsmath,amsthm}
\newtheorem{%
  definition}{Definition}
\newtheorem{%
  remark}{Remark}
\newtheorem{%
  lemma}{Lemma}
\newtheorem{%
  theorem}{Theorem}
\newtheorem{%
  fact}{Fact}
\newtheorem{%
  example}{Example}
\newtheorem{%
  proposition}{Proposition}
\usepackage{graphics,tikz}
\usetikzlibrary{arrows,decorations,snakes,backgrounds,patterns,calc}
\usetikzlibrary{shapes.geometric}

\newcommand{\natra}{}
\newcommand{\labdraw}{\ardrawl}

\newcommand{\strt}[1][]{\strut}
\newcommand{\maybe}[1][]{}

\newcommand{\mathrotate}[3][]{%
  \rotatebox[#1]{#2}{\ensuremath{#3}}%
}
  \newcommand{\poDia}{%
    \rotatebox[origin=r]{180}{\reflectbox{%
        \begin{tikzpicture}[baseline={(l.base)}]
          \tikzstyle{every node}+=[inner sep=0pt,outer sep=0pt]
      \node (l) at (0,0) {\vphantom{lg}};
      \node (r) at (1em,0) {\vphantom{lg}};    
      \begin{scope}[,shorten >=.5pt,shorten <=.5pt]
      \draw[very thin,->] (l.south) -- (l.north);
      \draw[very thin,->] (r.south) -- (r.north);
      \draw[very thin,->] (l.north) -- (r.north);
      \draw[very thin,->] (l.south) -- (r.south);
    \end{scope}
    \POC[.2]{r.north}{-135}
    \end{tikzpicture}%
  }}%
  }%
  
  \newcommand{\poDiaF}{\raisebox{.15ex}{\rotatebox[origin=r]{180}{\reflectbox{\poDia}}}}

  \newcommand{\poSqrF}[4]{%
    \ensuremath{{}_{\smaller{#1}}^{\smaller{#3}}\mkern2mu\poDiaF\mkern3mu{}^{\smaller{#4}}_{\smaller{#2}}}%
  }

  \newcommand{\pbDia}{%
    \raisebox{.4ex}{\rotatebox[origin=c]{180}{\scalebox{1.1}{\ensuremath{%
        {\smaller{{\mathrm{\kern-.1 em}\uparrow\mathrm{\kern-.1 em}}}}{%
          \makebox[0pt][l]{\raisebox{-.6ex}{\scalebox{.7}{\ensuremath{\urcorner}}}}%
        }{\vphantom{\raisebox{-.2ex}{g}}}^{\to}_{\to}{\smaller{{\mathrm{\kern-.1 em}\uparrow\mathrm{\kern-.1 em}}}}%
    }}}}%
  }%

  \newcommand{\MpbDia}{%
    \raisebox{.4ex}{\rotatebox[origin=c]{180}{\scalebox{1.1}{\ensuremath{%
        {\smaller{{\mathrm{\kern-.1 em}\mathrotate{90}{\hookrightarrow}}}}{%
          \makebox[0pt][l]{\raisebox{-.6ex}{\scalebox{.7}{\!\ensuremath{\urcorner}}}}%
        }{\vphantom{\raisebox{-.2ex}{g}}}^{\to}_{\to}{\smaller{{\mathrm{\kern-.05 em}\mathrotate{90}{\hookrightarrow}\mathrm{\kern-.1 em}}}}%
    }}}}%
  }%

  \newcommand{\pbcDia}{%
    \raisebox{.4ex}{\rotatebox[origin=c]{180}{\scalebox{1.1}{\ensuremath{%
        {\smaller{{\mathrm{\kern-.1 em}\uparrow\mathrm{\kern-.1 em}}}}{%
           \makebox[0pt][l]{\raisebox{-.6ex}{\scalebox{.7}[.6]{\ensuremath{\urcorner}}}}%
           \makebox[0pt][l]{\raisebox{-.6ex}{\scalebox{.7}[.8]{\ensuremath{\urcorner}}}}%
        }{\vphantom{\raisebox{-.2ex}{g}}}^{\to}_{\to}{\smaller{{\mathrm{\kern-.1 em}\uparrow\mathrm{\kern-.1 em}}}}%
    }}}}%
  }%

  \newcommand{\xMpoDia}{%
    \raisebox{.2ex}{\reflectbox{\scalebox{1.15}{\ensuremath{%
          {\smaller{\mathrotate{-90}{\hookrightarrow}}\kern-.05em}{\makebox[0pt][l]{\raisebox{-.6ex}{\scalebox{1}[.85]{\kern.05em\ensuremath{\urcorner}}}}}{\vphantom{\raisebox{-.2ex}{g}}}^{\gets}_{\scalebox{1.2}[1]{\makebox[0pt][l]{\ensuremath{\scriptstyle \gets}}}}{}%
    }}}}%
  }%
  \newcommand{\MpoDia}{%
    \raisebox{.2ex}{\reflectbox{\scalebox{1.15}{\ensuremath{%
          {\smaller{\mathrotate{-90}{\hookrightarrow}}\kern-.05em}{\makebox[0pt][l]{\raisebox{-.6ex}{\scalebox{1}[.85]{\kern.05em\ensuremath{\urcorner}}}}}{\vphantom{\raisebox{-.2ex}{g}}}^{\gets}_{\scalebox{1.2}[1]{\makebox[0pt][l]{\ensuremath{\scriptstyle \gets}}}}{\kern-.05em\smaller{\mathrotate{-90}{\hookrightarrow}}}%
    }}}}%
  }%

\newcommand{\squrflip}[5][r]{%
  \ifthenelse{\equal{#1}{r}}%
  {\SQURflip{#2}{#3}{#4}{#5}}%
  {\ensuremath{\mathreflect{\SQURflip[\mathreflect]{#2}{#3}{#4}{#5}}}}%
}

\newcommand{\SQURflip}[5][]{\ensuremath{{{}^{#1{#2}}}{}^{\to#1{#4}\smash{{}_{\searrow}}}_{{}^{\searrow}#1{#3}\to#1{#5}}}}

\newcommand{\rsqurflip}[5][r]{%
  \ifthenelse{\equal{#1}{r}}%
  {%
    \begin{tikzpicture}[very thin,baseline={([yshift=-3pt]D.north)}]
      \tikzstyle{every node}=[inner sep = 1.5pt]
      \node (A) {$\scriptscriptstyle #4$};
      \node[anchor=north] (D) at ([xshift=-.3ex]A.south) {$\scriptscriptstyle #3$};
      \draw[->] (A.west) -- ++(-1.3ex,0) coordinate (Beast);
      \node[anchor=east] (B) at (Beast) {$\scriptscriptstyle #2$};
      \draw[<-] (D.east) -- ++(1.3ex,0) coordinate (Cwest);
      \node[anchor=west] (C) at (Cwest) {$\scriptscriptstyle #5$};
      \draw[->] (B) -- (D);
      \draw[->] (A) -- (C);
    \end{tikzpicture}%
  }%
  {%
    \begin{tikzpicture}[very thin,baseline={([yshift=-3pt]D.north)},xscale=-1]
      \tikzstyle{every node}=[inner sep = 1.5pt]
      \node (A) {$\scriptscriptstyle #4$};
      \node[anchor=north] (D) at ([xshift=-.3ex]A.south) {$\scriptscriptstyle #3$};
      \draw[->] (A.west) -- ++(-1.3ex,0) coordinate (Beast);
      \node[anchor=east] (B) at (Beast) {$\scriptscriptstyle #2$};
      \draw[<-] (D.east) -- ++(1.3ex,0) coordinate (Cwest);
      \node[anchor=west] (C) at (Cwest) {$\scriptscriptstyle #5$};
      \draw[->] (B) -- (D);
      \draw[->] (A) -- (C);
    \end{tikzpicture}%
  }%
}

\newcommand{\squr}[5][r]{%
  \ifthenelse{\equal{#1}{r}}%
  {\SQUR{#2}{#3}{#4}{#5}}%
  {\ensuremath{\mathreflect{\SQUR[\mathreflect]{#2}{#3}{#4}{#5}}}}%
}

\newcommand{\SQUR}[5][]{\ensuremath{{{}_{#1{#2}}}_{\to#1{#4}\smash{{}^{\nearrow}}}^{{}_{\nearrow}#1{#3}\to#1{#5}}}}

\newcommand{\rsqur}[5][r]{%
  \ifthenelse{\equal{#1}{r}}%
  {\rSQUR{#2}{#3}{#4}{#5}}%
  {\ensuremath{\mathreflect{\rSQUR[\mathreflect]{#2}{#3}{#4}{#5}}}}%
}

\newcommand{\rSQUR}[5][]{\ensuremath{{{}_{#1{#2}}}_{\gets#1{#4}\smash{{}^{\nearrow}}}^{{}_{\nearrow}#1{#3}\gets#1{#5}}}}

\newcommand{\smaller}[1]{\mathord{%
    \mathchoice%
    {{\textstyle#1}}%
    {{\scriptstyle#1}}%
    {{\scriptscriptstyle#1}}%
    {{\scriptscriptstyle#1}}%
  }%
}

\newcommand{\mtoX}[3]{%
  \ifthenelse{\equal{#3}{}}{%
    \mathrel{\mkern-1mu\begin{tikzpicture}[baseline={(ANCHOR.base)}]%
        \node (ANCHOR) [anchor=base]  at (0,0) {\vphantom{a}};
        \node (text) [anchor=base,inner sep =.5pt,outer sep = 0pt] at (0,.25ex) {$\scriptstyle #3$};
        \draw[-#2] (text.base west) ++(0,.33ex) -- +(.4em,0);
        \draw[#1-] (text.base west)  ++(0,.33ex) ++ (-.4em,0) -- +(.4em,0); 
      \end{tikzpicture}\mkern-1mu}%
  }%
  {%
    \mathrel{\mkern-1mu\begin{tikzpicture}[baseline={(ANCHOR.base)}]%
        \node (ANCHOR) [anchor=base]  at (0,0) {\vphantom{a}};
        \node (text) [anchor=base,inner sep =.5pt,outer sep = 0pt] at (0,.25ex) {$\scriptstyle #3$};
        \draw[-#2] (text.base east) ++(0,.33ex) -- +(.5em,0);
        \draw[#1-] (text.base west)  ++(0,.33ex) ++ (-.5em,0) -- +(.5em,0); 
      \end{tikzpicture}\mkern-1mu}%
  }%
}

\newcommand{\mTOX}[3]{%
  \ifthenelse{\equal{#3}{}}{%
    \mathrel{\mkern-2mu\begin{tikzpicture}[baseline={(ANCHOR.base)}]%
        \node (ANCHOR) [anchor=base]  at (0,0) {\vphantom{a}};
        \node (text) [anchor=base,inner sep =.5pt,outer sep = 0pt] at (0,.25ex) {$\scriptstyle #3$};
        \draw[double,-#2] (text.base west) ++(0,.33ex) -- +(.4em,0);
        \draw[double,#1-] (text.base west)  ++(0,.33ex) ++ (-.4em,0) -- +(.4em,0); 
      \end{tikzpicture}\mkern-2mu}%
  }%
  {%
    \mathrel{\mkern-2mu\begin{tikzpicture}[baseline={(ANCHOR.base)}]%
        \node (ANCHOR) [anchor=base]  at (0,0) {\vphantom{a}};
        \node (text) [anchor=base,inner sep =.5pt,outer sep = 0pt] at (0,.25ex) {$\scriptstyle #3$};
        \draw (text.base east) ++(0,.33ex)[double,-] --
        ++(.5em,0) coordinate (tmp);
        \draw (tmp) [thick,-#2] -- +(.15em,0);
        \draw[double,#1-] (text.base west)  ++(0,.33ex) ++ (-.5em,0) -- +(.5em,0); 
      \end{tikzpicture}\mkern-2mu}%
  }%
}

\newcommand{\dsquare}{\Box\!\Box}

\renewcommand{\to}[1][]{\mtoX{}{>}{#1}}
\renewcommand{\gets}[1][]{\mtoX{<}{}{#1}}

\renewcommand{\hookleftarrow}[1][]{\mtoX{<}{<}{#1}}
\renewcommand{\hookrightarrow}[1][]{\mtoX{>}{>}{#1}}

\newcommand{\mgets}{\hookleftarrow}
\newcommand{\mto}{\hookrightarrow}
\newcommand{\cto}{\to\Delta}
\newcommand{\tra}[2][ ]{%
  \ifthenelse{\equal{#1}{ }}%
  {\trax{#2}}%
  {\trax{\langle #2,#1\rangle}}%
}
\newcommand{\trax}[1]{\!\mapstochar\,\mTOX{}{>}{#1}}

\makeatletter
\newcommand{\tikzslant}[1]{%
   \tikz[baseline=(N.base)]%
   \pgfsys@transformcm{1}{0}{0.22}{1}{0pt}{0pt}%
   \node[inner sep=0pt] (N) {#1};%
}
\newcommand{\slanttikz}[1]{%
   \tikz[baseline=(N.base)]%
   \pgfsys@transformcm{1}{0}{-0.22}{1}{0pt}{0pt}%
   \node[inner sep=0pt] (N) {#1};%
}

\makeatother

\usepackage[T1]{fontenc}
\newcommand{\CAT}[1]{
  {\ensuremath{\mathbb{\uppercase{#1}}}}}

\newcommand{\AR}[1]{
  {\ensuremath{\mathrm{ar}(\mathbb{#1})}}}

\newcommand{\OB}[1]{\objects{(\CAT{#1})}}
\newcommand{\objects}{\ensuremath{\mathrm{ob}}}

\newcommand{\ardrawL}{\ardrawl}

\newcommand{\coloncolon}{\fatsemi}

\newcommand{\arc}[3][--]{%
  \begin{scope}[]
    \begin{scope}[line width=0pt]
      \draw[draw=none] (#2) #1 coordinate[pos=.5](bla)  (#3) ;            
    \end{scope}
    \draw[>=open triangle 45,->,shorten >=-3.5pt,fill=white] (#2) #1 (bla); 
    \draw[shorten <=3.5pt] (bla) #1 (#3); 
  \end{scope}%
}

\newcommand{\arcC}[6][.5]{%
  \begin{scope}[]
    \tikzstyle{every node}=[]
    \begin{scope}[line width=0pt]
      \draw[draw=none] (#2) ..controls +(#4) and  +(#5)..  node{} coordinate[pos=#1](bla)  (#3) ;
    \end{scope}
    \draw[shorten >=3.5pt] (#3) ..controls +([scale=.6]#5) and +(#6:10pt).. (bla); 
    \draw[>=open triangle 45,->,shorten >=-3.5pt] (#2) .. controls +([scale=.6]#4) and +(180+#6:10pt) .. (bla); 
   \end{scope}%
}

\DeclareMathOperator{\col}{\mathrm{col}}
\DeclareMathOperator{\dom}{\mathrm{dom}}
\DeclareMathOperator{\cdm}{\mathrm{cdm}}

\newcommand{\outp}[3][0]{%
  \node
  (#2) 
  [scale=.4,draw,dart,shape border rotate=0+#1]
  at 
  (#3) {};        
}
\newcommand{\inp}[3][0]{%
  \node 
  [scale=.5,draw,kite,shape border rotate=90+#1]
  (#2) 
  at 
  (#3) {~};        
}    

\newcommand{\mytextcolor}{black}
\newcommand{\vtx}[3][]{%
  \ifthenelse{\equal{#1}{}}%
  {\begin{scope}[outer sep=0pt,shape=circle,inner sep=.5pt,thin]
    \tikzstyle{every node}+=[draw]
    \node (#2) at (#3) {\color{\mytextcolor}$\scriptstyle\kern0.02em {#2}\vphantom{\otimes{lg}}$} ;
  \end{scope}}%
  {\begin{scope}[outer sep=0pt,shape=circle,inner sep=.5pt,thin]
    \tikzstyle{every node}+=[draw]
    \node (#1) at (#3) {\color{\mytextcolor}$\scriptstyle\kern0.02em {#2}\vphantom{\otimes{lg}}$} ;
  \end{scope}}%
}

\usepackage{comment}
\includecomment{comment}

\begin{document}
\renewcommand{\marginpar}[2][]{}

\newcommand{\finitary}{finitary\xspace}
\newcommand{\finite}{finite\xspace}
\newcommand{\bb}{irreducible object\xspace}
\newcommand{\bbs}{\bb{}s\xspace}
\newcommand{\Bb}{Irreducible object\xspace}
\newcommand{\Bbs}{\Bb{}s\xspace}
\newcommand{\mono}{mono\xspace}
\newcommand{\monos}{monos\xspace}
\newcommand{\Mono}{Mono\xspace}
\newcommand{\Monos}{Monos\xspace}

\newcommand{\decomposition}{decomposition\xspace}

\maketitle
\begin{abstract}
  Inspired by decomposition problems in
  rule-based formalisms in Computational Systems Biology
  and recent work on compositionality in graph transformation, 
  this paper proposes to use \emph{arbitrary} colimits to ``deconstruct''
  models of reactions in which states are represented as objects 
  of adhesive categories.
  The fundamental problem is the decomposition 
  of complex reactions of large states
  into simpler reactions of smaller states. 

  The paper defines the \emph{local decomposition problem}
  for transformations. 
  To solve this problem means to  ``reconstruct''
  a given transformation
  as the  colimit of  ``smaller'' ones
  where the shape of the colimit and the decomposition of the source
  object of the transformation are fixed in advance. 
  The first result is the soundness of colimit decomposition for arbitrary
  double pushout transformations in any category, 
  which roughly means that several ``local''  transformations can be
  combined into a single ``global'' one. 
  Moreover, 
  a solution for a certain class of local decomposition
  problems is given, which generalizes and clarifies 
  recent work on compositionality in graph transformation.

\end{abstract}

\section*{Introduction}
\label{sec:introduction}
Compositional methods for the synthesis and analysis of computational
systems remain a fruitful research topic
with potential applications in practice. 
Though compositionality is most clearly exhibited in
semantics for process calculi where structural operational semantics
(\textsc{sos}) can be found in its ``pure'' form, a~slightly broader
perspective is appropriate to make use of the fundamental ideas of
\textsc{sos} in interdisciplinary research.

The first source of inspiration of the present paper is the
$\kappa$-calculus~\cite{DBLP:journals/tcs/DanosL04},
 which is an influential modelling framework in
Computational Systems Biology. The $\kappa$-calculus allows to give
abstract, formal descriptions of biological systems that can be used to
explain the reaction (rate) of complex systems, so-called
\emph{complexes}, in terms of the reaction (rate) of each of its
subsystems, which are called \emph{partial complexes}.
Leaving quantitative aspects as a topic for future research, 
we concentrate on a specific sub-problem, 
namely the ``purely structural'' decomposition of reactions.

In the $\kappa$-calculus,  
system states  are composed  of partial complexes
and they  have an intuitive, \emph{graphical} representation.
Hence, it is natural to investigate the decomposition of
(reactions of) system states using concepts from graph transformation. 
In its simplest form, 
the idea of composition of graph transformations 
is by means of coproducts.
Intuitively, 
the coproduct of two graphs models the assembly of two states put side
by side
and the two (sub-)states react independently of each other.
A well-known, related theorem about graph transformations
is the so-called Parallelism Theorem (see e.g.~\cite[Theorem~17]{DBLP:conf/gg/CorradiniMREHL97}).

A more general formalism of compositionality 
that is based on pushouts has been (re-)considered in~\cite{Rens10icalp}.
In the latter work, 
system states are constructed as pushouts in categories of graphs, 
and more generally states are modelled as objects in adhesive categories~\cite{lack:adhes}
(for related work in graph transformation on amalgamation and
distribution see~\cite{boehm1987amalgamation} 
and also~\cite[Section~6.2]{DBLP:conf/gg/CorradiniMREHL97}\nocite{DBLP:conf/gg/1997handbook}).
Roughly, 
\mbox{(de-)composition} via pushouts amounts
to ``splitting'' a state into two sub-states that are glued together
along a common interface. 
This  composition mechanism behaves  very much like the parallel composition in
name passing calculi;
in the composed process $P|Q$, 
the common interface of the two processes $P$ and $Q$
is essentially the intersection of their  free names.

In this paper, 
we shall remove the restriction to pushouts as a composition mechanism
and generalize the results of~\cite{Rens10icalp} 
from pushouts to (pullback stable) colimits of arbitrary shape. 
This considerably enlarges the set of available gluing patterns.
As a simple example, 
we can now equip each sub-state with several interfaces;
this would be appropriate for the  model of a cell in an organism
that is in direct contact with each of its neighbouring cells with some
part of its membrane;
each area of contact would be modelled by a different interface.

\paragraph{Content of the paper}
After reviewing some basic category theoretical concepts and the
definition of adhesive categories in Section~\ref{sec:preliminaries}, 
we begin Section~\ref{sec:gener-decomp-result} with the
``deconstruction'' of  models of system states;
more precisely, we explain in Section~\ref{sec:decomp-objects}
how suitably finite objects in adhesive
categories arise as the colimit of a diagram of ``atomic'' objects, 
namely irreducible objects in the sense of~\cite{BBHKlattice}.
We further give a short introduction to transformations of graph-like
structures in Section~\ref{sec:react-as-transf}
and review double pushout transformations~\cite{DBLP:conf/focs/EhrigPS73}\nocite{DBLP:conf/focs/FOCS14} 
as an transformation approach that -- in principle -- can be used in any category.
How the ``deconstruction'' of objects can be lifted to transformations
is described in Section~\ref{sec:decomp-transf}, 
and we conclude Section~\ref{sec:gener-decomp-result}
with a soundness result, 
which makes precise in what sense a family of ``local'' double pushout transformations
can be composed to a ``global'' transformation. 

The main problem, 
which is concerned with the decomposition
of a ``global'' transformation into a family of ``local'' ones, 
is addressed in Section~\ref{sec:decomp-probl}. 
We give a formal description of \emph{local decomposition problems}, 
which consist of a \emph{given} decomposition 
of a state (as a colimit of a certain shape)
and a rule that describes a possible reaction of the state;
to solve such a problem means to extend the decomposition 
of the state to a decomposition of the whole reaction (using colimits
of the same shape).
Section~\ref{sec:global-decomposition} presents a ``global'' solution,
which first constructs the whole transformation ``globally'';
a ``more local'' solution of the problem is possible
if we are given extra information that involve a generalization of the
accommodations of~\cite{Rens10icalp}. 
At the end of Section~\ref{sec:decomp-probl}, 
we present the \emph{Accommodated Completeness Theorem}, 
which gives a partial solution to the local decomposition problem
and generalizes the corresponding result of~\cite{Rens10icalp}. 
In Section~\ref{sec:conclusion}, we conclude with a summary of our
contributions, comment on related work, and point out directions for future research.



\section{Preliminaries}
\label{sec:preliminaries}
After fixing notation for categories, functors, natural
transformations and colimits, we recall the definition of adhesive
categories and give definitions of
further basic concepts; 
the standard reference for category theory is~\cite{MacLaneS:catwm}, 
\cite{PIERCE91} is a very accessible introduction.

Categories will be denoted by blackboard letters, $\CAT{C}, \CAT{D}$;
we write `$X \in \CAT{C}$' to express that $X$ is an object of $\CAT{C}$
and `$f \colon X \to Y$' in $\CAT{C}$ if $f$ is an arrow with domain $X$
and codomain $Y$.  
The identity on an object $X$ is  $\id_X \colon X \to X$, 
and the composition of two morphisms $f \colon X\to Y$ and $g \colon Y
\to Z$ is $g \circ f \colon X \to Z$. 

We usually use calligraphic letters to denote functors,
e.g.\ $\mathcal{F} \colon \CAT{C} \to \CAT{D}$ is a functor from the
category~$\CAT{C}$ 
to the category $\CAT{D}$.
Finally,  we use Greek letters 
to denote natural
transformations; hence the expression `$\tau \colon \mathcal{F}
\to \mathcal{G}$'
is used to indicate that $\tau$ is a natural transformation with 
domain $\mathcal{F}$ and codomain $\mathcal{G}$, 
where $\mathcal{F}, \mathcal{G}\colon \CAT{C} \to \CAT{D}$
are functors.
Given two categories $\CAT{C}, \CAT{D}$, 
the functor category that has  functors from $\CAT{C}$ to $\CAT{D}$
as objects and natural transformations between them 
as morphisms is denoted by $[\CAT{C},\CAT{D}]$. 

A central notion of this paper are colimits;
we follow the exposition in~\cite{MacLaneS:catwm}. 
\begin{figure}[htb]
  \centering
  $\mathcal{D}, \mathcal{E} \colon \CAT{J} \to \CAT{C}$\qquad
  $\begin{tikzpicture}[baseline={(D.base)},scale=1.4,yscale=.8]
    \mnode[D]{\mathcal{D}}{0,0}
    \mnode[colD]{\Delta D}{0,-1}
    \ardrawl{D}{colD}{swap}{\delta}
  \end{tikzpicture}$\qquad
  $\qquad
  \begin{tikzpicture}[baseline={(D.base)},scale=1.4,xscale=1.2,yscale=.8]
    \mnode[D]{\mathcal{D}}{0,0}
    \mnode[colD]{\Delta \col \mathcal{D}}{0,-1}
    \ardrawl{D}{colD}{swap}{\delta}
    \mnode[colC]{\Delta C}{1,-1}
    \ardrawl{D}{colC}{}{\gamma}
    \begin{scope}[dashed]
      \ardrawl{colD}{colC}{swap}{\Delta u}
    \end{scope}
  \end{tikzpicture}
  $\qquad
$
  \begin{tikzpicture}[baseline={(D.base)},scale=1.4,xscale=1.7,yscale=.8]
    \mnode[D]{\mathcal{D}}{0,0}
    \mnode[colD]{\Delta \col \mathcal{D}}{0,-1}
    \ardrawl{D}{colD}{swap}{\delta}
    \mnode[E]{\mathcal{E}}{1,0}
    \mnode[colC]{\Delta \col \mathcal{E}}{1,-1}
    \ardrawl{E}{colC}{}{\eta}
    \ardrawl{D}{E}{}{\tau}
    \begin{scope}[dashed]
      \ardrawl{colD}{colC}{swap}{\Delta \col \tau}
    \end{scope}
  \end{tikzpicture}
$
  \caption{Functors, cocones and colimits}
  \label{fig:fcc}
\end{figure}
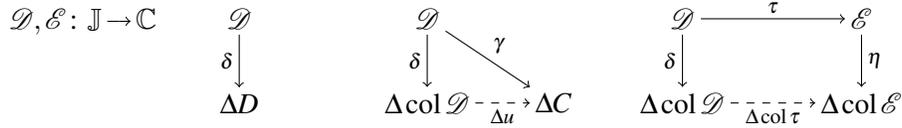
\begin{definition}[Cocones and colimits]
  \label{def:cocones-colimits}
  Let $\mathcal{D},\mathcal{E}\colon \CAT{J} \to \CAT{C}$ be functors.
  A \emph{cocone for $\mathcal{D}$} consists of an object $D \in
  \CAT{C}$ and a natural transformation $\delta \colon \mathcal{D} \to
  \Delta D$ 
  where $\Delta D \colon \CAT{J} \to \CAT{C}$ 
  is the functor that maps each object $i \in \CAT{J}$
  to $D$ and each morphism $f \colon i \to j$ in $\CAT{J}$
  to $\id_D$, the identity on $D$. 

  A cocone $\delta \colon \mathcal{D} \cto D$ 
  is a \emph{colimit}
  if for any other cocone $\gamma \colon \mathcal{D} \cto C$, 
  there is a unique morphism $u\colon D \to C$ in $\CAT{C}$ 
  such that $\gamma = \Delta u \circ \delta$ in $[\CAT{J}, \CAT{C}]$
  where $(\Delta u)_i = u$ for each $i\in \CAT{J}$. 
  The object $D$  of a colimit 
  $\delta \colon \mathcal{D} \cto D$ is 
  the \emph{colimit object} of $\mathcal{D}$
  and is often referred to as $\col \mathcal{D}$. 

  If $\delta \colon \mathcal{D} \cto D$ 
  and $\eta \colon \mathcal{E} \cto E$ 
  are  colimits 
  and moreover $\tau \colon \mathcal{D} \to \mathcal{E}$ 
  is a natural transformation, 
  then the unique morphism $t \colon D \to E$ 
  that satisfies $\eta \circ \tau = \Delta t \circ \delta$ 
  is denoted $\col \tau$.
\end{definition}
The notation for cocones and colimits is summarized in
Figure~\ref{fig:fcc}.
Colimits in the category of sets are particularly ``well-behaved'';
one relevant property is pullback stability, 
which we define together with cartesian transformations. 
\begin{definition}[Cartesian transformations and pullback stable colimits]
  \label{def:pb-stability-of-colimits}
  Let $\CAT{C},\CAT{J}$ be categories, 
  and let $\mathcal{K}, \mathcal{L}\colon \CAT{J} \to \CAT{C}$ 
  be functors.
  Further let 
  $\alpha \colon \mathcal{K} \to \mathcal{L}$
  be a natural transformation;
  it is \emph{cartesian}, if all naturality squares are pullbacks, 
  i.e.\ for all morphisms $f \colon i \to j$ in~$\CAT{J}$, 
  the span $\mathcal{K}_j 
  \gets [\mathcal{K}_f]
  \mathcal{K}_i 
  \to [\alpha_i] 
  \mathcal{L}_i$ 
  is a pullback of 
  $\mathcal{K}_j 
  \to [{\alpha_j}]
  \mathcal{L}_j 
  \gets[\mathcal{L}_f] 
  \mathcal{L}_i$.
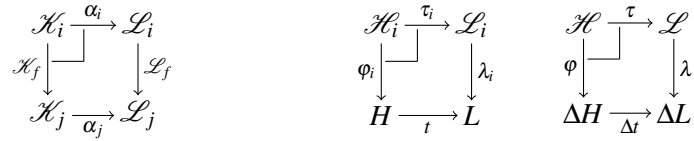
  \begin{displaymath}
    \begin{tikzpicture}[scale=1.2,baseline={(Ki.base)}]
      \mnode[Ki]{\mathcal{K}_i}{0,1}
      \mnode[Kj]{\mathcal{K}_j}{0,0}
      \ardrawl{Ki}{Kj}{swap}{\mathcal{K}_f}
      \mnode[Li]{\mathcal{L}_i}{1,1}
      \mnode[Lj]{\mathcal{L}_j}{1,0}
      \ardrawl{Li}{Lj}{}{\mathcal{L}_f}
      \ardrawl{Ki}{Li}{}{\alpha_i}
      \ardrawl{Kj}{Lj}{swap}{\alpha_j}
      \POC{Ki}{-45}
    \end{tikzpicture}\qquad \qquad \qquad 
    \begin{tikzpicture}[scale=1.2,baseline={(Ki.base)}]
      \mnode[Ki]{\mathcal{H}_i}{0,1}
      \mnode[Kj]{H}{0,0}
      \ardrawl{Ki}{Kj}{swap}{\varphi_i}
      \mnode[Li]{\mathcal{L}_i}{1,1}
      \mnode[Lj]{L}{1,0}
      \ardrawl{Li}{Lj}{}{\lambda_i}
      \ardrawl{Ki}{Li}{}{\tau_i}
      \ardrawl{Kj}{Lj}{swap}{t}
      \POC{Ki}{-45}
    \end{tikzpicture}\qquad
    \begin{tikzpicture}[scale=1.2,baseline={(Ki.base)}]
      \mnode[Ki]{\mathcal{H}}{0,1}
      \mnode[Kj]{\Delta H}{0,0}
      \ardrawl{Ki}{Kj}{swap}{\varphi}
      \mnode[Li]{\mathcal{L}}{1,1}
      \mnode[Lj]{\Delta L}{1,0}
      \ardrawl{Li}{Lj}{}{\lambda}
      \ardrawl{Ki}{Li}{}{\tau}
      \ardrawl{Kj}{Lj}{swap}{\Delta t}
      \POC{Ki}{-45}
    \end{tikzpicture}
  \end{displaymath}
  Let  $\lambda \colon \mathcal{L} \cto L$ be a colimit of
  $\mathcal{L}$;
  it is \emph{pullback stable}
  if for any cartesian transformation $\tau \colon \mathcal{H} \to
  \mathcal{L}$
  and every cocone $\varphi \colon \mathcal{H} \cto H$
  with a morphism $t \colon H \to L$ 
  such that the middle diagram in the above display is a pullback for all
  $i \in \CAT{J}$, 
  the cocone $\varphi \colon \mathcal{H} \cto H$ is actually a colimit
  of $\mathcal{H}$;
  in this situation, we say that $\varphi$ is the pullback of
  $\lambda$ along $t$
  (since the right one of the above diagrams is a pullback in
  $[\CAT{J},\CAT{C}]$). 
\end{definition}

The final definition on  basic category theoretical notions
concerns \monos, which are usually denoted using a ``tailed'' arrow $\mto$\,;
very roughly, \monos are a generalization of inclusions (of sets).
\begin{definition}[Subobjects and proper \monos]
  Let $\CAT{C}$ be a category, 
  and let $m \colon M \mto X$ and $n \colon N \mto X$  be \monos in $\CAT{C}$.
  Then \emph{$m$ is included in $n$}, 
  written $m \sqsubseteq n$, 
  if there exists a (mono-)morphism $f \colon M \to N$ such that $m = n \circ
  f$. 
  The \emph{subobject (represented by) $m$}
  is the collection
  \begin{displaymath}
    [m] = \Bigl\{
    m' \colon M' \mto X
    \bigm |
    \text{ $m'$ a \mono such that }
    m \sqsubseteq m'    \text{    and  } m' \sqsubseteq m
    \Bigr\};
  \end{displaymath}
  moreover, $[m] \sqsubseteq [n]$ if $m \sqsubseteq n$.
  A \emph{subobject of $X$}
  is a subobject $[m]$ such that the \mono $m$
  has $X$ as codomain, i.e.\ $m \colon M \mto X$.
  The \emph{subobject poset of $X$}, written $\Sub(X)$, 
  consists of all subobjects of $X$ and is ordered by the inclusion
  relation $\sqsubseteq$. 

 A \mono $m \colon M \mto X$ in $\CAT{C}$ is
 \emph{proper} if $\id_X \not \sqsubseteq m$, 
 and
 an object $X \in \CAT{C}$ \emph{has a proper
  subobject}
if there exist a proper \mono $n \colon N \mto X$ 
(which then represents a \emph{proper subobject} of $X$).   
\end{definition}

Systems states will be modelled as objects in adhesive categories, 
which have been introduced in~\cite{lack:adhes}
and are an established generalization of
(categories of) structures that are intuitively ``graph-like''.
\begin{definition}[Adhesive category]
  \label{def:adh-cate:2010}
  A category is \emph{adhesive} if
  \begin{itemize}
  \item it has pullbacks;
  \item it has pushouts along \monos;
    i.e.\ a pushout of a span $B \gets[f] A \hookrightarrow[m]C$ 
    exists if $m$ is a \mono; and
  \item 
    pushouts along \monos yield  \emph{Van Kampen \textsc{(vk)}
      squares},
    i.e.\ 
    if  ${B \to[n]D \gets[g] C}$ is the pushout 
    of $B \gets[f] A \hookrightarrow [m] C$ 
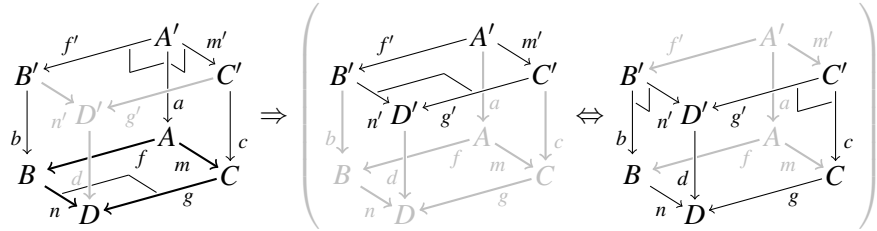
    \begin{center}
      \ensuremath{\displaystyle    %
  \!\!\begin{tikzpicture}[baseline=(current bounding box.west),scale=1.35]
      \tikzstyle{every node}=[circle,inner sep=0pt]
      \mnode{B}{-1,0,0} ;
      \node (C) at (1,0,0) {$C$} ;
      \node (A) at (0,0,-1) {$A$} ;
      \node (D) at (0,0,1) {$D$} ;
      \node (B') at (-1,1,0) {$B'$} ;
      \node (C') at (1,1,0) {$C'$} ;
      \node (A') at (0,1,-1) {$A'$} ;
      \begin{mycolor}{lightgray}
        \node (D') at (0,1,1) {$\boldsymbol{D'}$} ;
      \end{mycolor}
      \POCS[.5] D {-1,0,-1} {+1,0,-1} 
      \POCS {A'} {-1,0,+1} {0,-1,0}
      \POCS {A'} {+1,0,+1} {0,-1,0}
      \begin{scope}[thick]
        \ardrawL A B {below,very near start}{ f};
      \end{scope}
      \begin{scope}[thick]
        \ardrawL A C {below left,midway}{m\,};           
      \end{scope}
      \ardrawL {A'} {A} {right,near end}{a} ;
      \ardrawL{B'} {B} {left,near end}{b} ;      
      \ardrawL {C'} {C} {right,near end}{\,c};            
      \ardrawL {A'} {B'} {above,near end}{f'};
      \ardrawL {A'} {C'} {above right,midway}{\,m'};
      \begin{scope}[lightgray,thick]
        \ardrawL {B'} {D'} {below left, pos=.9}{\boldsymbol{n'\!\!}^{\vphantom l}} ;
      \end{scope}
      \begin{scope}[thick]
        \ardrawL {B} {D} {below left,near end}{n\,};
      \end{scope}
      \begin{scope}[lightgray,thick]
        \ardrawLd {D'} {D} {left,pos=.68}{\boldsymbol d} 
        \ardrawLd {C'} {D'} {below,near end}{\boldsymbol{g'}}       
      \end{scope}
      \begin{scope}[thick]
        \ardrawLd {C} {D} {below,near start}{g};
      \end{scope}
    \end{tikzpicture}
    \Rightarrow
    \color{lightgray}
    \left(
      \color{black}
      \begin{tikzpicture}[baseline=(current bounding box.west),scale=1.35]
        \tikzstyle{every node}=[circle,inner sep=0pt]
        \begin{scope}[lightgray,thick]
          \mnode[B]{\boldsymbol B}{-1,0,0} ;
          \node (C) at (1,0,0) {$\boldsymbol C$} ;
          \node (A) at (0,0,-1) {$\boldsymbol A$} ;
          \node (D) at (0,0,1) {$\boldsymbol D$} ;
        \end{scope}
        \node (B') at (-1,1,0) {$B'$} ;
        \node (C') at (1,1,0) {$C'$} ;
        \node (A') at (0,1,-1) {$A'$} ;
        \node (D') at (0,1,1) {$D'$} ;
        \POCS[.5] {D'} {-1,0,-1} {+1,0,-1} 
        \begin{scope}[lightgray,thick]
          \ardrawL A B {below,very near start}{\boldsymbol{ f}};
          \ardrawL A C {below left,midway}{\boldsymbol m\,}; 
          \ardrawL {A'} {A} {right,near end}{\boldsymbol a} ;
          \ardrawL{B'} {B} {left,near end}{\boldsymbol b} ;      
          \ardrawL {C'} {C} {right,near end}{\,\boldsymbol c};            
        \end{scope}
        \ardrawL {A'} {B'} {above,near end}{f'};
        \ardrawL {A'} {C'} {above right,midway}{\,m'};
        \ardrawL {B'} {D'} {below left, pos=.9}{{n'\!\!}^{\vphantom l}} ;      
        \begin{scope}[lightgray,thick]
          \ardrawL {B} {D} {below left,near end}{\boldsymbol n\,};      
          \ardrawLd {D'} {D} {left,pos=.68}{\boldsymbol d} ;
        \end{scope}
        \ardrawLd {C'} {D'} {below,near end}{{g'}} ;      
        \begin{scope}[lightgray,thick]
          \ardrawLd {C} {D} {below,near start}{\boldsymbol{g}}; 
        \end{scope}
      \end{tikzpicture}
      \Leftrightarrow
      \begin{tikzpicture}[baseline=(current bounding box.west),scale=1.35]
        \tikzstyle{every node}=[circle,inner sep=0pt]
        \mnode{B}{-1,0,0} ;
        \node (C) at (1,0,0) {$C$} ;
        \begin{scope}[lightgray,thick]
          \node (A) at (0,0,-1) {$\boldsymbol A$} ;
        \end{scope}
        \node (D) at (0,0,1) {$D$} ;
        \node (B') at (-1,1,0) {$B'$} ;
        \node (C') at (1,1,0) {$C'$} ;
        \begin{scope}[lightgray,thick]
          \node (A') at (0,1,-1) {$\boldsymbol{A'}$} ;
        \end{scope}
        \node (D') at (0,1,1) {$D'$} ;
        \POCS {C'} {-1,0,+1} {0,-1,0}
        \POCS {B'} {+1,0,+1} {0,-1,0}
        \begin{scope}[lightgray,thick]
          \ardrawL A B {below,very near start}{\boldsymbol{ f}}
          \ardrawL A C {below left,midway}{\boldsymbol m\,} 
          \ardrawL {A'} {A} {right,near end}{\boldsymbol a} 
        \end{scope}
        \ardrawL{B'} {B} {left,near end}{b}       
        \ardrawL {C'} {C} {right,near end}{\,c}            
        \begin{scope}[lightgray,thick]
          \ardrawL {A'} {B'} {above,near end}{\boldsymbol{f'}}
          \ardrawL {A'} {C'} {above right,midway}{\,\boldsymbol{m'}}
        \end{scope}
        \ardrawL {B'} {D'} {below left,pos=.9}{{n'\!\!}^{\vphantom l}} 
        \ardrawL {B} {D} {below left,near end}{n\,}     
        \ardrawLd {D'} {D} {left,pos=.68}{d} 
        \ardrawLd {C'} {D'} {below,near end}{g'}       
        \ardrawLd {C} {D} {below,near start}{g} 
      \end{tikzpicture}
      \color{lightgray}\right)\color{black}\!}%
      
    \end{center}
    then in each commutative cube over this square
    as shown above on the left,
    which has  pullbacks as back faces,
  the top~face is a pushout  
  if and only if
  the front faces are pullbacks (as illustrated above on the right).
  \end{itemize}
\end{definition}
The archetypal example of an adhesive category 
is the category of graphs, i.e.\ the functor category
$[{\cdot}\rightrightarrows{\cdot}, \CAT{SET}]$
where $\CAT{SET}$ is the usual category of sets and functions. 
One common property of the category of graphs and adhesive
categories is the fact that subobject posets are actually distributive
lattices.

\section{Colimits for composition and decomposition}
\label{sec:gener-decomp-result}
Composition of General Systems by means of colimits has been proposed by Goguen~\cite{goguen1973categorical}.
Based on this idea, we shall ``reconstruct''  objects 
as  colimits of their (primitive) constituents.
We illustrate the idea for the simplest type of colimits, 
namely  binary coproducts,
which generalize disjoint unions of sets to objects in arbitrary categories.
The question whether an object is the coproduct of two ``components''
amounts to asking whether the object is \emph{connected};
in fact, a graph $G$ is connected if and only if any ``reconstruction'' of
$G$ as a coproduct $G \cong H_1 + H_2$ is ``trivial''
in the sense that $H_1$ or $H_2$ is the empty graph.

The next more general type of colimit are fibred coproducts, 
i.e.\ pushouts. 
In fact, composition  by pushouts of pairs of \monos
is the basic construction principle that is considered in recent work
on compositionality of graph transformation~\cite{Rens10icalp}. 
Roughly, to decompose an object in this manner means to find 
an ``interface subobject'' such that the original object can be 
glued together from two (other) subobjects that overlap on the
``interface subobject''.

\begin{figure}[htb]
  \centering
  \begin{tikzpicture}[baseline={(v.base)}]
    \node [matrix,fill=white!80!black,column
    sep=1em,row sep=1em]{
    \vtx{v}{0,0}
    \vtx{w}{1,-.5}
    \vtx{u}{-1,-.5}
    \arc{v}{u}
    \arc{w}{v}\\
  };
  \end{tikzpicture}
  \quad  \quad  $\leadsto$ \quad  \quad
  \begin{tikzpicture}[scale=1.2, baseline={(v1.base)}]
    \node
    (v)
    [matrix,fill=white!80!black,column
    sep=1em,row sep=1em]
    at (0,0)
    {
    \vtx[v0]{v}{0,0}
    \\
  };
    \node
    (vu)
    [matrix,fill=white!80!black,column
    sep=1em,row sep=1em]
    at (-1,-1)
    {
    \vtx[v1]{v}{0,0}
     \vtx{u}{-1,-.5}
     \arc{v1}{u}
    \\
  };
  \node
  (wv)
  [matrix,fill=white!80!black,column
    sep=1em,row sep=1em]
    at (1,-1)
    {
    \vtx[v2]{v}{0,0}
     \vtx{w}{1,-.5}
     \arc{w}{v2}
    \\
  };
  \ardrawl{v}{vu}{}{}
  \ardrawl{v}{wv}{}{}
  \begin{scope}[densely dotted]
    \draw[bend left] (v0) to (v1);
    \draw[bend right] (v0) to (v2);
  \end{scope}
  \end{tikzpicture}
  \caption{Pushout decomposition of a connected graph}
  \label{fig:POdecomp}
\end{figure}
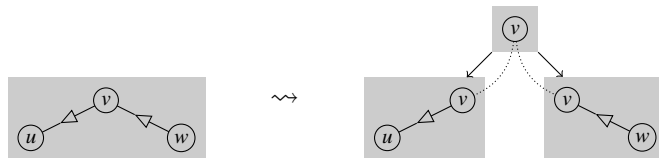

A simple example of a pushout decomposition of a connected graph is given in
Figure~\ref{fig:POdecomp}. 
The ``interface'' is just the node $v$, 
and the two edges are glued together at $v$;
this is very much like the parallel composition $u.v.0
|v.w.0$ of the  two $\textsc{ccs}$ processes $u.v.0$ and $v.w.0$. 
However, we are not restricted to have (sets of) nodes as ``interfaces''.

\begin{remark}
To avoid confusion, 
it is important to mention that the above notion of ``interface subobject''
seems not to be directly connected
to the notion of interface that is used in techniques for the
derivation of labelled transition systems for reactive systems~\cite{DBLP:journals/njc/SassoneS03}
and graph transformation~\cite{EH06}.
We hope that in the preset paper, 
this source of confusion is less problematic
as we use arbitrary colimit shapes and not just assemblies of
``interfaces''.
Moreover, we do not consider the possibility to change the shapes 
of diagrams as it is done for the case of graph transformation in~\cite{GabiPhD}, 
because we want to avoid the problem that colimits cannot be constructed componentwise. 
\end{remark}

Next, based on results of~\cite{BBHKlattice}, 
we  introduce and illustrate a canonical decomposition procedure.
The transformation framework that will be used later is introduced 
in Section~\ref{sec:react-as-transf}
before we describe how the decomposition procedure can be lifted
from objects to transformations 
(Section~\ref{sec:decomp-transf});
a soundness theorem for decomposition of transformations
forms the end of this section.

\subsection{Decomposition of objects}
\label{sec:decomp-objects}
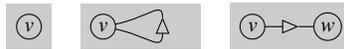
\begin{figure}[htb]
  \centering
  
  \begin{tikzpicture}[scale=1.5,baseline={(v.base)}]
    \node [matrix,fill=white!80!black,column
    sep=1em,row sep=1em]
    at (-1,0) {
      \vtx{v}{0,0}
      \\
      };
    \end{tikzpicture}\quad
    \begin{tikzpicture}[scale=1.5,baseline={(v.base)}]
    \node [matrix,fill=white!80!black,column
    sep=1em,row sep=1em]
    at (0,0) 
    {
      \vtx{v}{0,0}
      \begin{scope}[overlay]
        \arcC{v}{v}{1,-.2}{1,.2}{90}        
      \end{scope}
      \phantom{\vtx{x}{1,0}}
      \\
      };
    \end{tikzpicture}\quad
    \begin{tikzpicture}[scale=1.5,baseline={(v.base)}]
      \node [matrix,fill=white!80!black,column
    sep=1em,row sep=1em]
    at (1,0) 
    {
      \vtx{v}{0,0}
      \vtx{w}{1,0}
      \arc{v}{w}
      \\
      };
  \end{tikzpicture}
  \caption{The three \bbs in the category of graphs}
  \label{fig:building-blocks}
\end{figure}

To ensure that ``concrete'' \decomposition{}s of states and reactions do exist, 
we assume that each state can be modelled as an object with finitely many primitive constituents,
which can be thought of as atoms, molecules or peptides 
 -- depending on whether we use  metaphors from 
Physics, Chemistry, or Microbiology.
If we use graphs as models, 
a canonical set of \emph{\bbs{}} consists of nodes, loops, and
edges (with distinct source and target nodes); these ``atoms'' of
graphs are illustrated in Figure~\ref{fig:building-blocks}.

In \textsc{uml} models and informal graphical  system specifications, 
a large family of different ``graph-like'' structures is used; 
typically, graph-like structures can either be encoded as graphs
or they are instances of adhesive categories~\cite{lack:adhes}, 
which are an established framework that captures a large variety of
structures that are intuitively  ``graph-like''. 
Given an object (with finitely many subobjects) in an arbitrary adhesive category,
its irreducible components can be characterized lattice theoretically
as (the sources of) its irreducible subobjects; 
here, we shall use the following equivalent Definition~\cite{BBHKlattice}.
\begin{definition}[\Bb]
  \label{def:bb}
  Let $\CAT{C}$ be  an adhesive category
  and let $X \in \CAT{C}$ be an object;
  the object $X$ is \emph{(subobject) \finite{}} if the lattice $\Sub(X)$ is finite.

  Let $Y \in \CAT{C}$ be a \finite object; 
  it is an \emph{\bb{}} if 
  \begin{enumerate}
  \item 
    it has a proper subobject;
  \item
    in every 
    pushout $U \mto[u] Y\mgets[v] V$
    of a span of \monos $U \mgets W \mto V$, 
    at most one of $u$ and $v$ represents a proper subobject of $Y$.
  \end{enumerate}
  A~\bb $Y$ is an \emph{\bb{}} or {\emph{an irreducible component} of
    $X$} if there exists a \mono $y \colon Y \mto X$.
\end{definition}
\noindent 
The first requirement of this definition ensures that $Y$ is
``non-trivial'';
in the case of graphs, it rules out the empty graph as an~\bb.

Similar to the ``deconstruction'' of a given lattice element as a join
of irreducible ones, 
\finite objects in adhesive categories arise as the colimit of
irreducible components~\cite[Corollary~1]{BBHKlattice}.
The relevant consequence in the context of the present paper is the
following fact, 
which actually goes beyond Birkhoff's representation theorem for
distributive lattices. 
\begin{fact}[Object decomposition]
  \label{fact:object-decomposition} 
  Let $\CAT{C}$ be an adhesive category, and let $X \in \CAT{C}$ be a \finite object
  with a proper subobject. 

  There exists a diagram $\mathcal{X} \colon \CAT{J} \to \CAT{C}$
  with a pullback stable colimit $\xi \colon \mathcal{X} \cto
  X$
  such that for each $j \in \CAT{J}$,
  $\xi_{j} \colon \mathcal{X}_j \mto X$ 
  is a \mono and $\mathcal{X}_j$  either is an \bb 
  or it does not have a proper subobject.
\end{fact}

\begin{figure}[htb]
  \centering
  \begin{tikzpicture}[baseline={(x.base)}]
    \node [matrix,fill=white!80!black,column
    sep=1em,row sep=1em,]{
    \vtx{x}{0,0}
    \vtx{w}{.5,.5}
    \vtx{u}{-.5,.5}
    \arc{u}{w}
    \arc{u}{x}
    \arc{w}{x}
    \arcC{w}{w}{.5,-.5}{.5,.5}{90}
    \\
  };
  \end{tikzpicture}
  \qquad \raisebox{4ex}{$\leadsto$} \qquad
  \begin{tikzpicture}[scale=1.3,xscale=1.2,baseline={(xx.base)}
    ]
    \node
    (nu)
    [matrix,fill=white!80!black,column
    sep=1em,row sep=1em,]
    at (-1,0)
    {%
    \vtx[u']{u}{0,0}%
    \\%
  };
    \node
    (nx)
    [matrix,fill=white!80!black,column
    sep=1em,row sep=1em,]
    at (0,0)
    {
    \vtx[x']{x}{0,0}
    \\
  };  
    \node
    (nw)
    [matrix,fill=white!80!black,column
    sep=1em,row sep=1em,]
    at (1,0)
    {
    \vtx[w']{w}{0,0}%

    \\%
  };  

    \node
    (ux)
    [matrix,fill=white!80!black,column
    sep=1em,row sep=1em,]
    at (-1,-1)
    {
      \vtx[xx]{x}{0,-.5}
      \vtx{u}{-.5,0}
     \arc{u}{xx}
    \\
  };  
      \node
    (uw)
    [matrix,fill=white!80!black,column
    sep=1em,row sep=1em,]
    at (0,-1)
    {
      \vtx{w}{.5,0}
      \vtx{u}{-.5,0}
      \arc{u}{w}
    \\
  };  
    \node
    [matrix,fill=white!80!black,column
    sep=1em,row sep=1em,]
    (wx)
    at (1,-1)
    {
      \vtx{x}{0,-.5}
      \vtx{w}{.5,0}
     \arc{w}{x}
    \\
  };  
    \node
    (ww)
    [matrix,fill=white!80!black,column
    sep=1em,row sep=1em,]
    at (2,-1)
    {
      \vtx{w}{.5,0}
     \arcC{w}{w}{.5,-.5}{.5,.5}{90}
    \\
  };  
  \foreach \u/\v in {nu/ux,nu/uw,%
    nw/uw,nw/wx,nw/ww,%
    nx/ux,nx/wx}
  {\ardrawld{\u}{\v}{}{}
    };
  \end{tikzpicture}
  \caption{Canonical colimit decomposition of a graph}
  \label{fig:canonicalDecomp}
\end{figure}
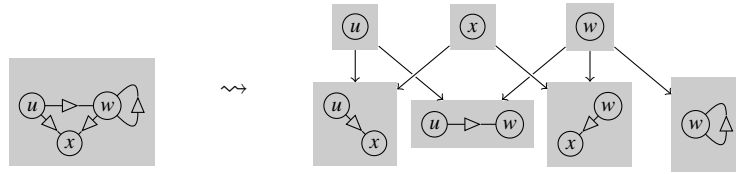

\begin{remark}
In fact, 
there are \emph{canonical} decompositions of finite objects.
In Fact~\ref{fact:object-decomposition}, 
the role of the category $\CAT{J}$ is played
by the sub-lattice of $\Sub(X)$ 
that contains all irreducibles and the bottom element;
the functor $\mathcal{X}$ chooses a representative $m \colon M \mto X$ of each subobject
$[m] \in \CAT{J} \subseteq \Sub(X)$ and maps it to the domain $M$, 
which  either is an irreducible object or it does not have a proper
subobject. 
\end{remark}
\noindent
 An example of a canonical decomposition is the graph  on
the left in Figure~\ref{fig:canonicalDecomp},
which is the colimit object of the diagram of irreducible objects on the
right in Figure~\ref{fig:canonicalDecomp} where all morphisms are inclusions.

Summarizing,
we assume objects to be finite to be able to  ``reconstruct'' each object as a pullback stable
colimit of irreducible objects. 
In Section~\ref{sec:decomp-transf}, we shall describe an analogous
notion of decomposition for reactions
if they are modelled by one of the more common (graph) transformation
approaches, 
which are covered by the modelling framework for reactions of systems
that we introduce next.

\subsection{Reactions as double square  transformations}
\label{sec:react-as-transf}
To describe reactions and their effects on complex system states, 
we shall use rule-based models. 
As a  typical example of a rule, consider  a pair of inclusions $L \supseteq K
\subseteq R$ of sets, graphs or other graph-like objects.
The reaction of a state that is represented by some object~$X$ 
corresponds to an \emph{application} of this rule
if the left hand side  $L$  is included in $X$, i.e.\ $L \subseteq X$, 
and the state after the reaction is modelled by an object~$Z$
that is obtained from $X$ by the following construction:
first, remove from $X$ everything that is  covered by $L$ but not by
$K$ and call the intermediate result $Y$;
second, adjoin to $Y$ everything in $R$ that is not covered
by $K$. 
In $\CAT{Set}$, we have the following double square diagram. 
\begin{displaymath}
 \begin{tikzpicture}[scale=1.7,yscale=.7,baseline={(0,-.5)}]
    \mnode{L}{-1,0}
    \mnode{K}{0,0}
    \mnode{R}{1,0}
    \ardrawL{K}{L}{above}{\supseteq}
    \ardrawL{K}{R}{above}{\subseteq}
    \begin{scope}[shift={(0,-1)}]
      {
        \mnode[A]{X}{-1,0}
      \ardrawL{L}{A}{left}{\rotatebox{-90}{$\scriptstyle \subseteq$}}
      }
     \end{scope}
  \end{tikzpicture}\qquad\qquad
  \begin{tikzpicture}[scale=1.7,yscale=.7,baseline={(0,-.5)}]
    \mnode{L}{-1,0}
    \mnode{K}{0,0}
    \mnode{R}{1,0}
    \ardrawL{K}{L}{above}{\supseteq}
    \ardrawL{K}{R}{above}{\subseteq}
    \begin{scope}[shift={(0,-1)}]
      {
        \mnode[A]{X}{-1,0}
      \ardrawL{L}{A}{left}{\rotatebox{-90}{$\scriptstyle \subseteq$}}
      }
      {
        \mnode[D]{X \setminus (L\setminus K)\makebox[0pt][l]{\color{gray}$\boldsymbol{{}=:Y}$}}{0,0}
        \ardrawL{D}{A}{below}{\supseteq}
        \ardrawL{K}{D}{left}{\rotatebox{-90}{$\scriptstyle \subseteq$}}
      }
    \end{scope}
  \end{tikzpicture}\qquad\qquad
  \begin{tikzpicture}[scale=1.7,yscale=.7,baseline={(0,-.5)}]
    \mnode{L}{-1,0}
    \mnode{K}{0,0}
    \mnode{R}{1,0}
    \ardrawL{K}{L}{above}{\supseteq}
    \ardrawL{K}{R}{above}{\subseteq}
    \begin{scope}[shift={(0,-1)}]
      {
        \mnode[A]{X}{-1,0}
      \ardrawL{L}{A}{left}{\rotatebox{-90}{$\scriptstyle \subseteq$}}
      }
            {
        \mnode[D]{Y}
        {0,0}
        \ardrawL{D}{A}{below}{\supseteq}
        \ardrawL{K}{D}{left}{\rotatebox{-90}{$\scriptstyle \subseteq$}}
      }
      {
      \mnode[B]{Y \uplus (R\setminus K)}{1,0}
      \ardrawL{D}{B}{below}{\subseteq}                 
      \ardrawL{R}{B}{right}{\rotatebox{-90}{$\scriptstyle\subseteq$}}
      }
    \end{scope}
  \end{tikzpicture}
\end{displaymath}
After a formalization of rules, 
we continue with examples, 
and define transformations by means of double square diagrams in any category, 
which will give us a fairly general modelling framework for reactions. 
\begin{definition}[Rule]
  A rule is an arbitrary span of morphisms $L \gets[a] K \to[b] R$, 
  where $L$ is the \emph{left hand side \textsc{(lhs)}}, 
  $R$ is the \emph{right hand side \textsc{(rhs)}}, 
  and $K$ is the \emph{context} or \emph{interface} of the rule.
\end{definition}
Rules are usually ranged over by letters $p$ and $q$.\footnote{%
  The letter `p' is taken from the word `production';
  in fact, rules are meant to generalize productions of  Chomsky grammars. 
}
Intuitively, 
in any rule, 
the left morphism describes deletion (and cloning) actions and, dually,  
the right one specifies addition of new entities (and merging of existing ones).
As an illustrative example of a rule
consider the graphical encoding of reaction in our \emph{Mini Solos Calculus}.

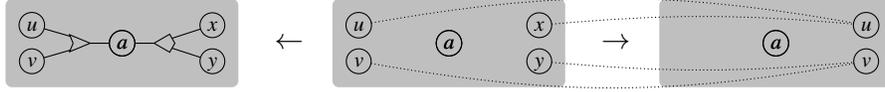
\begin{figure}[htb]
  \centering
   \tikzstyle{every picture}=+[background rectangle/.style=
     {rounded corners=.5ex,draw=none,fill=lightgray}, show background rectangle]
      \begin{tikzpicture}[scale=.6,baseline={(a.base)},remember picture]
      \begin{scope}
        \outp{i}{0,0}
        \vtx{a}{1,0}
        \draw (i.tip) --(a);
        \vtx{u}{-1,.4}
        \vtx{v}{-1,-.4}
        \draw (i.south west) -- (v);
        \draw (i.north west) -- (u);
      \end{scope}
      \begin{scope}
        \inp[180]{i}{2,0}
        \vtx{a}{1,0}
        \draw (i.lower vertex) --(a);
        \vtx{x}{3,.4}
        \vtx{y}{3,-.4}

        \draw (i.north east) -- (x);
        \draw (i.south east) -- (y);
      \end{scope}
    \end{tikzpicture}
    \quad$\leftarrow$\quad
      \begin{tikzpicture}[scale=.6,baseline={(a.base)},remember picture]
      \begin{scope}
        \vtx{a}{1,0}
        \vtx{u}{-1,.4}
        \vtx{v}{-1,-.4}
      \end{scope}
      \begin{scope}
        \vtx{a}{1,0}
        \vtx{x}{3,.4}
        \vtx{y}{3,-.4}
      \end{scope}
    \end{tikzpicture}
    \quad$\rightarrow$\quad
      \begin{tikzpicture}[scale=.6,baseline={(a.base)},remember picture]
      \begin{scope}
        \vtx{a}{1,0}
        \phantom{\vtx[u'']{u}{-1,.4}}
        \phantom{\vtx[u'']{v}{-1,-.4}}
      \end{scope}
      \begin{scope}
        \vtx{a}{1,0}
        \vtx[u']{u}{3,.4}
        \vtx[v']{v}{3,-.4}
      \end{scope}
      \begin{scope}[densely dotted]
       \draw[overlay,bend left=10] (u) to (u');
       \draw[overlay,bend right=10] (v) to (v');
       \draw[overlay,bend left=5] (x) to (u');
       \draw[overlay,bend right=5] (y) to (v');
      \end{scope}
    \end{tikzpicture}
    %
  \caption{The single rule for reaction of solos}
  \label{fig:solosGTS}
\end{figure}

\begin{example}[Mini Solos Calculus]
  We define a 
  fragment of the calculus of explicit fusions~\cite{xi-calc},
  which in turn is closely related to the solos calculus~\cite{solos}. 
  \begin{displaymath}
    {
      \colorbox{white!80!black}{$P ::= a!uv \bigm | a?xy \bigm | P|P \bigm | 0$}
    }
    \qquad\qquad
    {
      \colorbox{white!80!black}{$P | a!uv | a?xy ~ \to ~ P[x/u,y/v]$}
    }
  \end{displaymath}
  A process is essentially a multiset of \emph{solos}, 
  which come in two variants, namely \emph{outputs} $a!uv$ 
  and \emph{inputs}  $a?xy$
  where $a,u,v,x,y$ are elements of a countable set of names.

  To keep the example short, 
  processes are considered as elements of a commutative monoid $(\CAT{P}, |, 0)$, 
  all names are free, 
  and the operational semantics is given ``globally''. 
  In any process of the form  $P | a!uv | a?xy$, 
  the last two solos synchronize on $a$ and ``fuse'' the pairs $(u,x)$ 
  and $(v,y)$, 
  which is achieved by simultaneous substitution of $u$ for $x$ and
  $v$ for $y$ in $P$
  (see~\cite{xi-calc} for an \textsc{sos} style presentation, 
  where \emph{explicit fusions} and a  sophisticated
  structural congruence are used).

    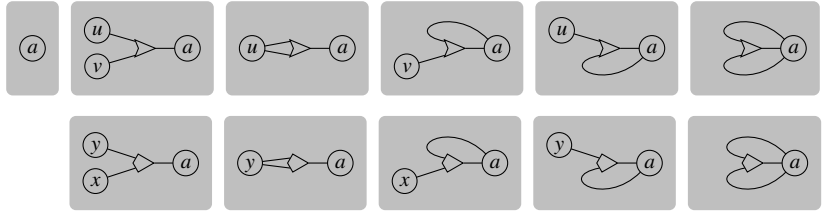
\begin{figure}[htb]
  \centering
   \tikzstyle{every picture}=+[background rectangle/.style=
     {rounded corners=.5ex,draw=none,fill=lightgray}, show background rectangle]
     \begin{tikzpicture}[scale=.6,baseline={(a.base)}]
      \begin{scope}
        \vtx{a}{0,0}
      \end{scope}
        \draw[draw=none] (a.north) -- +(0,1.5ex) node[pos=1]{~};
        \draw[draw=none] (a.south) -- +(0,-1.5ex) node[pos=1]{~};
    \end{tikzpicture}\,
    \begin{tikzpicture}[scale=.6,baseline={(a.base)}]
      \begin{scope}
        \outp{i}{0,0}
        \vtx{a}{1,0}
        \draw (i.tip) --(a);
        \vtx{u}{-1,.4}
        \vtx{v}{-1,-.4}
        \draw (i.south west) -- (v);
        \draw (i.north west) -- (u);
      \end{scope}
        \draw[draw=none] (a.north) -- +(0,1.5ex) node[pos=1]{~};
        \draw[draw=none] (a.south) -- +(0,-1.5ex) node[pos=1]{~};
        \draw[draw=none] (i.west) -- +(-4ex,0) node[pos=1]{~};
    \end{tikzpicture}\,
    \begin{tikzpicture}[scale=.6,baseline={(a.base)}]
      \begin{scope}
        \outp{i}{0,0}
        \vtx{a}{1,0}
        \draw (i.tip) --(a);
        \vtx{u}{-1,0}
        \draw (i.south west) -- (u);
        \draw (i.north west) -- (u);
      \end{scope}
        \draw[draw=none] (a.north) -- +(0,1.5ex) node[pos=1]{~};
        \draw[draw=none] (a.south) -- +(0,-1.5ex) node[pos=1]{~};
        \draw[draw=none] (i.west) -- +(-4ex,0) node[pos=1]{~};
    \end{tikzpicture}\,
    \begin{tikzpicture}[scale=.6,baseline={(a.base)}]
      \begin{scope}
        \outp{i}{0,0}
        \vtx{a}{1,0}
        \draw (i.tip) --(a);
        \vtx{v}{-1,-.4}
        \draw (i.south west) -- (v);
        \draw[overlay] (i.north west) .. controls +(-1,.2) and +(-1,1) .. (a);
      \end{scope}
        \draw[draw=none] (a.north) -- +(0,1.5ex) node[pos=1]{~};
        \draw[draw=none] (a.south) -- +(0,-1.5ex) node[pos=1]{~};
        \draw[draw=none] (i.west) -- +(-4ex,0) node[pos=1]{~};
    \end{tikzpicture}\,
    \begin{tikzpicture}[scale=.6,baseline={(a.base)}]
      \begin{scope}
        \outp{i}{0,0}
        \vtx{a}{1,0}
        \draw (i.tip) --(a);
        \vtx{u}{-1,.4}
        \draw[overlay] (i.south west) .. controls +(-1,-.2) and +(-1,-1) .. (a);
        \draw (i.north west) -- (u);
      \end{scope}
        \draw[draw=none] (a.north) -- +(0,1.5ex) node[pos=1]{~};
        \draw[draw=none] (a.south) -- +(0,-1.5ex) node[pos=1]{~};
        \draw[draw=none] (i.west) -- +(-4ex,0) node[pos=1]{~};
    \end{tikzpicture}\,
    \begin{tikzpicture}[scale=.6,baseline={(a.base)}]
      \begin{scope}
        \outp{i}{0,0}
        \vtx{a}{1,0}
        \draw (i.tip) --(a);
        \draw[overlay] (i.south west) .. controls +(-1,-.2) and +(-1,-1) .. (a);
        \draw[overlay] (i.north west) .. controls +(-1,.2) and +(-1,1)
        .. (a);
        \draw[draw=none] (a.north) -- +(0,1.5ex) node[pos=1]{~};
        \draw[draw=none] (a.south) -- +(0,-1.5ex) node[pos=1]{~};
        \draw[draw=none] (i.west) -- +(-4ex,0) node[pos=1]{~};
      \end{scope}
    \end{tikzpicture}\\[1ex]
    \,\begin{tikzpicture}[background rectangle/.style={fill=white},scale=.6,baseline={(a.base)}]
      \begin{scope}
        \phantom{\vtx{a}{0,0}}
      \end{scope}
        \draw[draw=none] (a.north) -- +(0,1.5ex) node[pos=1]{~};
        \draw[draw=none] (a.south) -- +(0,-1.5ex) node[pos=1]{~};
    \end{tikzpicture}\,
    \begin{tikzpicture}[scale=.6,baseline={(a.base)}]
      \begin{scope}
        \inp{i}{0,0}
        \vtx{a}{1,0}
        \draw (i.lower vertex) --(a);
        \vtx{y}{-1,.4}
        \vtx{x}{-1,-.4}
        \draw (i.south west) -- (x);
        \draw (i.north west) -- (y);
      \end{scope}
        \draw[draw=none] (a.north) -- +(0,1.5ex) node[pos=1]{~};
        \draw[draw=none] (a.south) -- +(0,-1.5ex) node[pos=1]{~};
        \draw[draw=none] (i.west) -- +(-4ex,0) node[pos=1]{~};
    \end{tikzpicture}\,
    \begin{tikzpicture}[scale=.6,baseline={(a.base)}]
      \begin{scope}
        \inp{i}{0,0}
        \vtx{a}{1,0}
        \draw (i.lower vertex) --(a);
        \vtx{y}{-1,0}
        \draw (i.south west) -- (y);
        \draw (i.north west) -- (y);
      \end{scope}
        \draw[draw=none] (a.north) -- +(0,1.5ex) node[pos=1]{~};
        \draw[draw=none] (a.south) -- +(0,-1.5ex) node[pos=1]{~};
        \draw[draw=none] (i.west) -- +(-4ex,0) node[pos=1]{~};
    \end{tikzpicture}\,
    \begin{tikzpicture}[scale=.6,baseline={(a.base)}]
      \begin{scope}
        \inp{i}{0,0}
        \vtx{a}{1,0}
        \draw (i.lower vertex) --(a);
        \vtx{x}{-1,-.4}
        \draw (i.south west) -- (x);
        \draw[overlay] (i.north west) .. controls +(-1,.2) and +(-1,1) .. (a);
      \end{scope}
        \draw[draw=none] (a.north) -- +(0,1.5ex) node[pos=1]{~};
        \draw[draw=none] (a.south) -- +(0,-1.5ex) node[pos=1]{~};
        \draw[draw=none] (i.west) -- +(-4ex,0) node[pos=1]{~};
    \end{tikzpicture}\,
    \begin{tikzpicture}[scale=.6,baseline={(a.base)}]
      \begin{scope}
        \inp{i}{0,0}
        \vtx{a}{1,0}
        \draw (i.lower vertex) --(a);
        \vtx{y}{-1,.4}
        \draw[overlay] (i.south west) .. controls +(-1,-.2) and +(-1,-1) .. (a);
        \draw (i.north west) -- (y);
      \end{scope}
        \draw[draw=none] (a.north) -- +(0,1.5ex) node[pos=1]{~};
        \draw[draw=none] (a.south) -- +(0,-1.5ex) node[pos=1]{~};
        \draw[draw=none] (i.west) -- +(-4ex,0) node[pos=1]{~};
    \end{tikzpicture}\,
    \begin{tikzpicture}[scale=.6,baseline={(a.base)}]
      \begin{scope}
        \inp{i}{0,0}
        \vtx{a}{1,0}
        \draw (i.lower vertex) --(a);
        \draw[overlay] (i.south west) .. controls +(-1,-.2) and +(-1,-1) .. (a);
        \draw[overlay] (i.north west) .. controls +(-1,.2) and +(-1,1)
        .. (a);
        \draw[draw=none] (a.north) -- +(0,1.5ex) node[pos=1]{~};
        \draw[draw=none] (a.south) -- +(0,-1.5ex) node[pos=1]{~};
        \draw[draw=none] (i.west) -- +(-4ex,0) node[pos=1]{~};
      \end{scope}
    \end{tikzpicture}
  \caption{Irreducible objects  in the graphical encoding of the Mini
    Solos Calculus}
  \label{fig:solosGraphs}
\end{figure}
  The Mini Solos Calculus has a simple, precise encoding as a graph
  transformation system.
  According to Definition~\ref{def:bb},
  the irreducible objects are names
  and input and output solos 
  as shown   Figure~\ref{fig:solosGraphs}
  where darts represent outputs and kites represent inputs. 
  The single reaction rule is given in Figure~\ref{fig:solosGTS}. 
\end{example}

Independent of the choice of a particular transformation approach, 
a rule  is applied to an object $X$
by mapping (part of) its left hand side to $X$. 
In the more common approaches, 
namely double-, single- and sesqui-pushout 
rewriting~\cite{DBLP:conf/focs/EhrigPS73,DBLP:journals/tcs/Lowe93,DBLP:confggCorradiniHHK06},
an application of a rule to an object $X$ yields a transformed
object $Y$ provided that there exists a suitable double square diagram that has
the same shape as the one in Figure~\ref{fig:double-square-trans}.
    
  \begin{figure}[htb]
    \centering
    \begin{tikzpicture}[baseline={(bl)}]
      \mnode{L}{-1,1}
      \mnode{K}{0,1}
      \mnode{R}{1,1}
      \ardrawl{K}{L}{swap}{a}
      \ardrawl{K}{R}{}{b}
      
      \mnode{X}{-1,0}
      \mnode{Y}{0,0}
      \mnode{Z}{1,0}
      \ardrawl{Y}{X}{}{c\vphantom{d}}
      \ardrawl{Y}{Z}{swap}{d}
      \foreach \u/\v/\m/\s in {L/X/m/swap,K/Y/j/swap,R/Z/n/}
      \ardrawl{\u}{\v}{\s}{\m};
      \draw[draw=none] (R) -- (X) coordinate[pos=.5] (bl);
    \end{tikzpicture}\qquad
    \begin{tabular}[c]{rl}
      rule: & $q = (L \gets[a] K \to[b] R)$ \\
      match: & $L \to[m] X$\\
      transformation: & $X \tra[m]{q} Y$
    \end{tabular}
    \caption{Double square transformation}
    \label{fig:double-square-trans}
  \end{figure}
Such a diagram models the reaction of the state $X$, 
and different reaction types correspond to different rules. 
To avoid lengthy elaborations on the technical details of single-, double-
and sesqui-pushout transformation, 
we use the following definition of transformation.
\begin{definition}[Double square transformation]
  \label{def:double-square-trans}
  Let $\CAT{C}$ be any category. 
  A \emph{(double square) transformation approach (in $\CAT{C}$)}
  is a collection $\CAT{D}$ of commutative double square diagrams in $\CAT{C}$,
  i.e.\ each element of $\CAT{D}$ is a commutative diagram as shown in
  Figure~\ref{fig:double-square-trans}.%
  \footnote{%
    In other words, a transformation approach is essentially a full
    subcategory $\CAT{D}$
    of the functor category $[\dsquare, \CAT{C}]$ 
    where $\dsquare$ is the obvious category with six objects and
    fifteen morphisms -- including identities.
    }

  Let $\CAT{D}$ be a double square approach, 
  and let $q = (L \gets[a] K \to[b] R)$ be a rule. 
  A~morphism $m \colon L \to X$ is a \emph{match}
  if there exist a pair of composable morphisms 
  $X \gets[c] Y \gets [j] K$
  and a cospan $Y \to[d] Z \gets[n]R$ 
  such that the diagram in Figure~\ref{fig:double-square-trans} is
  a member of $\CAT{D}$;
  in such a situation,
  \emph{$q$ transforms $X$ to $Y$ (at the match~$m$)}, 
  and we express this by writing $X \tra[m]{q} Y$
  or just $X \tra{q} Y$.
  Sometimes, we refer to `$X \tra[m]{q} Y$' as a
  \emph{transformation}
  to implicitly refer to the double square diagram in Figure~\ref{fig:double-square-trans}.
\end{definition}

In the case of graphs, 
different requirements on the exact nature of the squares
correspond to different policies for  the implicit deletion of
dangling edges and the resolution of so-called conflicts of deletion
and preservation (see~\cite{DBLP:conf/gg/CorradiniMREHL97}). 
The most well-known instance is the double pushout approach to
rewriting~\cite{DBLP:conf/focs/EhrigPS73}, 
which is also the approach that is used in~\cite{Rens10icalp}.
\begin{definition}[Double pushout approach]
  \label{def:DPO}
  Let $\CAT{C}$ be any category. 

  \parpic[r]{
  \begin{tikzpicture}[baseline={(bl)}]
      \mnode{L}{-1,1}
      \mnode{K}{0,1}
      \mnode{R}{1,1}
      \ardrawl{K}{L}{swap}{a}
      \ardrawl{K}{R}{}{b}
      \mnode{X}{-1,0}
      \mnode{Y}{0,0}
      \mnode{Z}{1,0}
      \ardrawl{Y}{X}{}{c\vphantom{d}}
      \ardrawl{Y}{Z}{swap}{d}
      \foreach \u/\v/\m/\s in {L/X/m/swap,K/Y/j/swap,R/Z/n/}
      \ardrawl{\u}{\v}{\s}{\m};
      \draw[draw=none] (R) -- (X) coordinate[pos=.5] (bl);
      \POC{X}{45}
      \POC{Z}{135}
    \end{tikzpicture}    
}
The \emph{double pushout approach}, 
  denoted by $\CAT{DPO}$, 
  consists of all double square diagrams that comprise two pushout
  squares in $\CAT{C}$, 
  i.e.\ $\CAT{DPO}$ contains all diagrams of the form that is
  illustrated to the right;
  %
  the angles indicate that $L \to[m] X \gets[c]Y$ 
  is a pushout of $L \gets[a] K \to[j] Y$ 
  and similarly $Y \to[d] Z \gets[n] R$ is the pushout 
  of $Y \gets[j] K \to[b] R$. 
\end{definition}

\subsection{Decomposition of reactions}
\label{sec:decomp-transf}
If we use a double square approach to model reactions, 
we  can apply the notion of object decomposition from
Section~\ref{sec:decomp-objects}
to each object in a double square diagram 
and obtain a notion of decomposition for transformations.
If we restrict to double pushout rewriting, 
we can generalize the soundness theorem of~\cite{Rens10icalp},
which ensures that families of ``local'' transformations 
can be composed to obtain a ``global'' transformation;
the complementary  (partial) completeness theorem
of~\cite{Rens10icalp} will be discussed in
Section~\ref{sec:decomp-probl}. 

We start with a formal definition of decompositions 
of double square diagrams. 
\begin{definition}[Transformation decomposition]
  \label{def:trans-decomp}
  Let $\CAT{D}$ be a double square transformation approach in $\CAT{C}$
  and let $\CAT{J}$ be a category;
  moreover, 
  let the left one of the following diagrams be a member of $\CAT{D}$.
  \begin{displaymath}
    \begin{tikzpicture}[baseline={(bl)}]
      \mnode{L}{-1,1}
      \mnode{K}{0,1}
      \mnode{R}{1,1}
      \ardrawl{K}{L}{swap}{a}
      \ardrawl{K}{R}{}{b}
      
      \mnode{X}{-1,0}
      \mnode{Y}{0,0}
      \mnode{Z}{1,0}
      \ardrawl{Y}{X}{}{c\vphantom{d}}
      \ardrawl{Y}{Z}{swap}{d}
      \foreach \u/\v/\m/\s in {L/X/m/swap,K/Y/j/swap,R/Z/n/}
      \ardrawl{\u}{\v}{\s}{\m};
      \draw[draw=none] (R) -- (X) coordinate[pos=.5] (bl);
    \end{tikzpicture} \qquad \leadsto \qquad 
      \begin{tikzpicture}[baseline={(bl)}]
      \mnode[L]{\mathcal L}{-1,1}
      \mnode[K]{\mathcal K}{0,1}
      \mnode[R]{\mathcal R}{1,1}
      \ardrawl{K}{L}{swap}{\alpha }
      \ardrawl{K}{R}{}{\beta}
      
      \mnode[X]{\mathcal X}{-1,0}
      \mnode[Y]{\mathcal Y}{0,0}
      \mnode[Z]{\mathcal Z}{1,0}
      \ardrawl{Y}{X}{}{\gamma}
      \ardrawl{Y}{Z}{swap}{\delta}
      \foreach \u/\v/\m/\s in {L/X/\mu/swap,K/Y/\iota/swap,R/Z/\nu/}
      \ardrawl{\u}{\v}{\s}{\m};
      \draw[draw=none] (R) -- (X) coordinate[pos=.5] (bl);
    \end{tikzpicture} 
    \qquad
      \begin{tikzpicture}[baseline={(bl)}]
      \mnode[L]{\mathcal L_i}{-1,1}
      \mnode[K]{\mathcal K_i}{0,1}
      \mnode[R]{\mathcal R_i}{1,1}
      \ardrawl{K}{L}{swap}{\alpha_i }
      \ardrawl{K}{R}{}{\beta_i}
      
      \mnode[X]{\mathcal X_i}{-1,0}
      \mnode[Y]{\mathcal Y_i}{0,0}
      \mnode[Z]{\mathcal Z_i}{1,0}
      \ardrawl{Y}{X}{}{\gamma_i}
      \ardrawl{Y}{Z}{swap}{\delta_i}
      \foreach \u/\v/\m/\s in {L/X/\mu_i/swap,K/Y/\iota_i/swap,R/Z/\nu_i/}
      \ardrawl{\u}{\v}{\s}{\m};
      \draw[draw=none] (R) -- (X) coordinate[pos=.5] (bl);
    \end{tikzpicture} 
  \end{displaymath}
   A \emph{$\CAT{J}$-decomposition} of this diagram 
   is a commutative double square diagram in $[\CAT{J},\CAT{C}]$,
   as shown above in the middle, 
    such that 
    the right hand diagram in the above display 
    belongs to $\CAT{D}$
    for each $i \in \CAT{J}$, 
    and moreover the  diagram  in  the functor category
    $[\CAT{J},\CAT{C}]$
    that is shown in Figure~\ref{fig:decomposing-double-square-diagrams}
    \begin{figure}[htb]
      \centering
      \begin{tikzpicture}[baseline={(bl)},scale=1.7]
      \mnode[L]{\Delta L}{-1,1}
      \mnode[K]{\Delta K}{0,1}
      \mnode[R]{\Delta R}{1,1}
      \ardrawl{K}{L}{swap}{\Delta a}
      \ardrawl{K}{R}{}{\Delta b}

      \mnode[X]{\Delta X}{-1,0}
      \mnode[Y]{\Delta Y}{0,0}
      \mnode[Z]{\Delta Z}{1,0}
      \ardrawl{Y}{X}{}{\Delta c}
      \ardrawl{Y}{Z}{swap}{\Delta d}
      \foreach \u/\v/\m/\s in {L/X/\Delta m/swap,K/Y/\Delta j/swap,R/Z/\Delta n/}
      \ardrawl{\u}{\v}{\s}{\m};
      \draw[draw=none] (R) -- (X) coordinate[pos=.5] (bl);
      \begin{scope}[shift={(.75,-.75)},scale=2.3]
      \mnode[L']{\mathcal L}{-1,1}
      \mnode[K']{\mathcal K}{0,1}
      \mnode[R']{\mathcal R}{1,1}
      \ardrawl{K'}{L'}{swap}{\alpha }
      \ardrawl{K'}{R'}{}{\beta}
      
      \mnode[X']{\mathcal X}{-1,0}
      \mnode[Y']{\mathcal Y}{0,0}
      \mnode[Z']{\mathcal Z}{1,0}
      \ardrawl{Y'}{X'}{}{\gamma}
      \ardrawl{Y'}{Z'}{swap}{\delta}
      \foreach \u/\v/\m/\s in {L/X/\mu/swap,K/Y/\iota/swap,R/Z/\nu/}
      \ardrawld{\u'}{\v'}{\s}{\m};
      \end{scope}
      \foreach \u/\m/\s in {L/\lambda/,K/\kappa/swap,R/\rho/,X/\xi/swap,Y/\upsilon/,Z/\zeta/swap}
      \ardrawld{\u'}{\u}{\s}{\m};
    \end{tikzpicture}
      \caption{Colimit decomposition of a double square diagram}
      \label{fig:decomposing-double-square-diagrams}
    \end{figure}
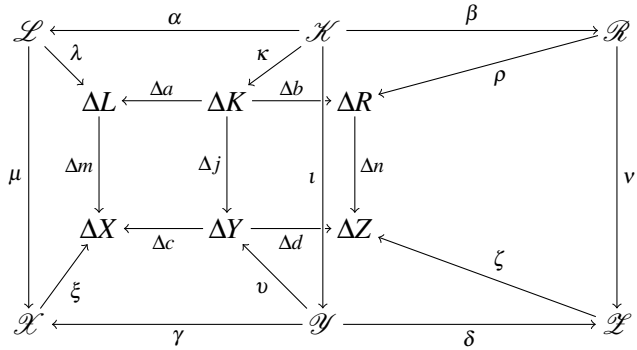
    commutes 
    where all cocones are colimits, 
    i.e.\ $\lambda, \kappa, \rho, \xi, \upsilon$, and $\zeta$ are colimits.
\end{definition}
Roughly, 
to decompose a double square diagram means to find a double square
diagram in the relevant functor category such that the original diagram
is induced by taking colimits of the six functors. 
The following example uses the category ${\cdot} \gets {\cdot} \to
{\cdot}$ 
for $\CAT{J}$ and $\CAT{C} = [{\cdot}\rightrightarrows{\cdot},
\CAT{SET}]$
is the category of graphs.
\begin{example}[Decomposition of a double pushout diagram]
  \label{ex:DPOdecomp}
  Consider the following small rule,
  which features the main actions that we have used 
  in the  Mini Solos Calculus;
  the rule deletes the edge from $x$ to $u$
  and fuses the two nodes $x$ and $z$, 
  which become $v$.
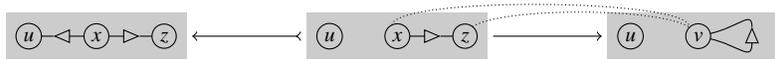
  \begin{displaymath}
     \begin{tikzpicture}[xscale=4]
       \node 
       [matrix,fill=white!80!black,column sep=1em,row sep=1em]
       (L)
       at (-1,0)
       {
         \vtx{u}{-.9,0}
         \vtx{x}{0,0}
         \vtx{z}{.9,0}
         \arc{x}{u}
         \arc{x}{z}
         \\
       };
       \node 
       [matrix,fill=white!80!black,column sep=1em,row sep=1em]
       (K)
       at (0,0)
       {
         \vtx{u}{-.9,0}
         \vtx{x}{0,0}
         \vtx{z}{.9,0}
         \arc{x}{z}
         \\
       };
       \node 
       [matrix,fill=white!80!black,column sep=1em,row sep=1em]
       (R)
       at (1,0)
       {
         \vtx{u}{-.9,0}
         \vtx[x']{v}{0,0}
         \phantom{\vtx[z']{z}{.9,0}}
         \begin{scope}[overlay]
           \arcC{x'}{x'}{.9,-.2}{.9,.2}{90}           
         \end{scope}
         \\
       };
       \begin{scope}[shorten >=2pt,shorten <=2pt,]
         \draw [->] (K) -- (R);
         \draw [>->] (K) -- (L);
       \end{scope}
       \begin{scope}[densely dotted]
         \draw[bend left=90] (x) to (x'.135) ;
         \draw[bend left=70] (z) to (x'.135) ;
       \end{scope}
     \end{tikzpicture}
  \end{displaymath}
  An application of this rule is shown at the top of
  Figure~\ref{fig:PO-decomposition}.
  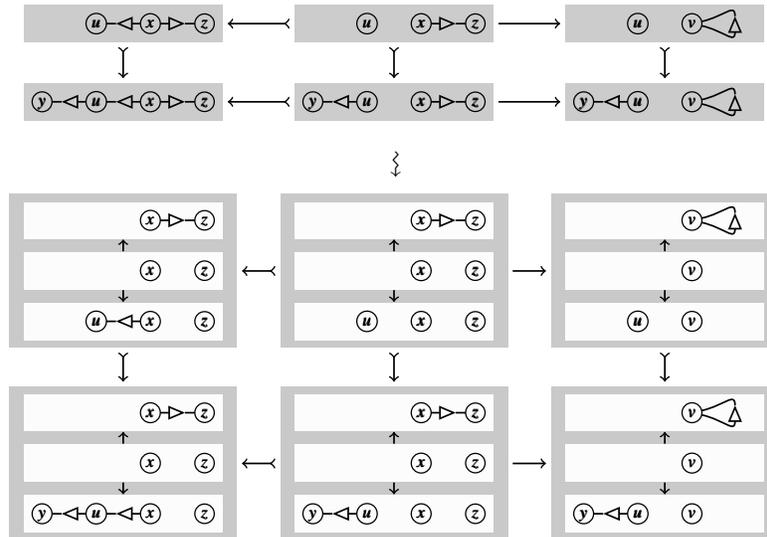
\begin{figure}[htb]
\renewcommand{\vtx}[3][]{%
  \ifthenelse{\equal{#1}{}}%
  {\begin{scope}[outer sep=0pt,shape=circle,inner sep=.5pt,semithick]
    \tikzstyle{every node}+=[draw]
    \node (#2) at (#3) {\color{\mytextcolor}$\scriptstyle\kern0.02em {\pmb{#2}}\vphantom{\otimes{lg}}$} ;
  \end{scope}}%
  {\begin{scope}[outer sep=0pt,shape=circle,inner sep=.5pt,semithick]
    \tikzstyle{every node}+=[draw]
    \node (#1) at (#3) {\color{\mytextcolor}$\scriptstyle\kern0.02em {\pmb{#2}}\vphantom{\otimes{lg}}$} ;
  \end{scope}}%
}
    \centering
    \scalebox{.8}{\begin{tikzpicture}[xscale=4.5,yscale=1.3,thick]
       \node 
       [matrix,thick,fill=white!80!black,column sep=1em,row sep=1em]
       (L)
       at (-1,0)
       {
         \vtx{u}{-.9,0}
         \vtx{x}{0,0}
         \vtx{z}{.9,0}
         \begin{scope}[shift={(-.9,0)}]
           \phantom{\vtx{y}{-.9,0}}
         \end{scope}
         \arc{x}{u}
         \arc{x}{z}
         \\
       };
       \node 
       [matrix,thick,fill=white!80!black,column sep=1em,row sep=1em]
       (K)
       at (0,0)
       {
         \vtx{u}{-.9,0}
         \vtx{x}{0,0}
         \vtx{z}{.9,0}
          \begin{scope}[shift={(-.9,0)}]
            \phantom{\vtx{y}{-.9,0}}
          \end{scope}
         \arc{x}{z}
         \\
       };
       \node 
       [matrix,thick,fill=white!80!black,column sep=1em,row sep=1em]
       (R)
       at (1,0)
       {
         \vtx{u}{-.9,0}
         \vtx[x']{v}{0,0}
         \phantom{\vtx[z']{z}{.9,0}}
         \begin{scope}[overlay]
           \arcC{x'}{x'}{.9,-.2}{.9,.2}{90}           
         \end{scope}
         \begin{scope}[shift={(-.9,0)}]
           \phantom{\vtx{y}{-.9,0}}
          \end{scope}
         \\
       };
       \node 
       [matrix,thick,fill=white!80!black,column sep=1em,row sep=1em]
       (X)
       at (-1,-1)
       {
         \vtx{u}{-.9,0}
         \vtx{x}{0,0}
         \vtx{z}{.9,0}
         \begin{scope}[shift={(-.9,0)}]
           \vtx{y}{-.9,0}           
         \end{scope}
         \arc{x}{u}
         \arc{x}{z}
         \arc{u}{y}
         \\
       };
       \node 
       [matrix,thick,fill=white!80!black,column sep=1em,row sep=1em]
       (Y)
       at (0,-1)
       {
         \vtx{u}{-.9,0}
         \vtx{x}{0,0}
         \vtx{z}{.9,0}
          \begin{scope}[shift={(-.9,0)}]
            \vtx{y}{-.9,0}           
          \end{scope}
         \arc{x}{z}
         \arc{u}{y}
         \\
       };
       \node 
       [matrix,thick,fill=white!80!black,column sep=1em,row sep=1em]
       (Z)
       at (1,-1)
       {
         \vtx{u}{-.9,0}
         \vtx[x']{v}{0,0}
         \phantom{\vtx[z']{z}{.9,0}}
         \begin{scope}[overlay]
           \arcC{x'}{x'}{.9,-.2}{.9,.2}{90}           
         \end{scope}
          \begin{scope}[shift={(-.9,0)}]
            \vtx{y}{-.9,0}           
          \end{scope}
         \arc{u}{y}
         \\
       };
       \begin{scope}[shorten >=2pt,shorten <=2pt,]
         \draw [->] (Y) -- (Z);
         \draw [>->] (Y) -- (X);
       \end{scope}
       \foreach \u/\v in {L/X,K/Y,R/Z}
       \draw[shorten >=2pt,shorten <=2pt,>->] (\u) -- (\v);
       \begin{scope}[shorten >=2pt,shorten <=2pt,]
         \draw [->] (K) -- (R);
         \draw [>->] (K) -- (L);
       \end{scope}
     \end{tikzpicture}}\\[-.5ex]
     \rotatebox{-90}{$\leadsto$}
     \\[1ex]
     \scalebox{.8}{\begin{tikzpicture}[xscale=4.5,yscale=3.2,thick]
       \node[fill=white!79!black]
       (L) at (-1,0) {%
         \parbox{.22\textwidth}{%
           \centering
           \tikz[remember picture]
           \node
           [matrix,thick,fill=white!99!black,column sep=1em,row sep=1em]
           (up)
           {
             \vtx{x}{0,0}            
              \begin{scope}[shift={(-.9,0)}]
                \phantom{\vtx{y}{-.9,0}}
              \end{scope}
              \vtx{z}{.9,0}
              \arc{x}{z}
             \\
           };%
           \\[1.1ex]
         \tikz[remember picture]
            \node
            [matrix,thick,fill=white!99!black,column sep=1em,row sep=1em]
            (interface)
            {
              \phantom{\vtx{u}{-.9,0}}
              \vtx{x}{0,0}
              \vtx{z}{.9,0}
              \phantom{%
                \begin{scope}[shift={(-.9,0)}]
                  \phantom{\vtx{y}{-.9,0}}
                \end{scope}
                \arc{x}{u}
                \arc{x}{z}
              }
              \\
            };
         \\[1.1ex]
         \tikz[remember picture]
         \node
         [matrix,thick,fill=white!99!black,column sep=1em,row sep=1em]
         (down)
         {
           \vtx{x}{0,0}
           \vtx{u}{-.9,0}
           \begin{scope}[shift={(-.9,0)}]
             \phantom{\vtx{y}{-.9,0}}
           \end{scope}
           \arc{x}{u}
           \vtx{z}{.9,0}
           \\
         };%
         \tikz[remember picture, overlay] \draw[->,thick](interface.north) --
         (up.south);%
         \tikz[remember picture, overlay] \draw[->,thick](interface.south) --
         (down.north);%
          }
        };
       \node[fill=white!79!black]
       (K) at (0,0)
       {%
         \parbox{.22\textwidth}{%
           \centering
           \tikz[remember picture]
           \node
           [matrix,thick,fill=white!99!black,column sep=1em,row sep=1em]
           (up)
           {
             \vtx{x}{0,0}            
              \begin{scope}[shift={(-.9,0)}]
                \phantom{\vtx{y}{-.9,0}}
              \end{scope}
              \vtx{z}{.9,0}
              \arc{x}{z}
             \\
           };%
           \\[1.1ex]
         \tikz[remember picture]
            \node
            [matrix,thick,fill=white!99!black,column sep=1em,row sep=1em]
            (interface)
            {
              \phantom{\vtx{u}{-.9,0}}
              \vtx{x}{0,0}
              \vtx{z}{.9,0}
              \phantom{%
                \begin{scope}[shift={(-.9,0)}]
                  \phantom{\vtx{y}{-.9,0}}
                \end{scope}
                \arc{x}{u}
                \arc{x}{z}
              }
              \\
            };
         \\[1.1ex]
         \tikz[remember picture]
         \node
         [matrix,thick,fill=white!99!black,column sep=1em,row sep=1em]
         (down)
         {
           \vtx{x}{0,0}
           \vtx{u}{-.9,0}
           \begin{scope}[shift={(-.9,0)}]
             \phantom{\vtx{y}{-.9,0}}
           \end{scope}
           \vtx{z}{.9,0}
           \\
         };%
         \tikz[remember picture, overlay] \draw[->,thick](interface.north) --
         (up.south);%
         \tikz[remember picture, overlay] \draw[->,thick](interface.south) --
         (down.north);%
          }
        };
        \node[fill=white!79!black]
       (R) at (1,0)
       {%
         \parbox{.22\textwidth}{%
           \centering
           \tikz[remember picture]
           \node
           [matrix,thick,fill=white!99!black,column sep=1em,row sep=1em]
           (up)
           {
             \vtx{v}{0,0}            
              \begin{scope}[shift={(-.9,0)}]
                \phantom{\vtx{y}{-.9,0}}
              \end{scope}
              \phantom{\vtx{z}{.9,0}}
              \begin{scope}[overlay]
                \arcC{v}{v}{.9,-.2}{.9,.2}{90}           
              \end{scope}
             \\
           };%
           \\[1.1ex]
         \tikz[remember picture]
            \node
            [matrix,thick,fill=white!99!black,column sep=1em,row sep=1em]
            (interface)
            {
              \phantom{\vtx{u}{-.9,0}}
              \vtx{v}{0,0}
              \phantom{\vtx{z}{.9,0}}
              \phantom{%
                \begin{scope}[shift={(-.9,0)}]
                  \phantom{\vtx{y}{-.9,0}}
                \end{scope}
              }
              \\
            };
         \\[1.1ex]
         \tikz[remember picture]
         \node
         [matrix,thick,fill=white!99!black,column sep=1em,row sep=1em]
         (down)
         {
           \vtx{v}{0,0}
           \vtx{u}{-.9,0}
           \begin{scope}[shift={(-.9,0)}]
             \phantom{\vtx{y}{-.9,0}}
           \end{scope}
           \phantom{\vtx{z}{.9,0}}
           \\
         };%
         \tikz[remember picture, overlay] \draw[->,thick](interface.north) --
         (up.south);%
         \tikz[remember picture, overlay] \draw[->,thick](interface.south) --
         (down.north);%
          }
        };
       \begin{scope}[shorten >=2pt,shorten <=2pt,]
         \draw [->] (K) -- (R);
         \draw [>->] (K) -- (L);
       \end{scope}
       \node[fill=white!79!black]
       (X) at (-1,-1) {%
         \parbox{.22\textwidth}{%
           \centering
           \tikz[remember picture]
           \node
           [matrix,thick,fill=white!99!black,column sep=1em,row sep=1em]
           (up)
           {
             \vtx{x}{0,0}            
              \begin{scope}[shift={(-.9,0)}]
                \phantom{\vtx{y}{-.9,0}}
              \end{scope}
              \vtx{z}{.9,0}
              \arc{x}{z}
             \\
           };%
           \\[1.1ex]
         \tikz[remember picture]
            \node
            [matrix,thick,fill=white!99!black,column sep=1em,row sep=1em]
            (interface)
            {
              \phantom{\vtx{u}{-.9,0}}
              \vtx{x}{0,0}
              \vtx{z}{.9,0}
              \phantom{%
                \begin{scope}[shift={(-.9,0)}]
                  \phantom{\vtx{y}{-.9,0}}
                \end{scope}
                \arc{x}{u}
                \arc{x}{z}
              }
              \\
            };
         \\[1.1ex]
         \tikz[remember picture]
         \node
         [matrix,thick,fill=white!99!black,column sep=1em,row sep=1em]
         (down)
         {
           \vtx{x}{0,0}
           \vtx{u}{-.9,0}
           \begin{scope}[shift={(-.9,0)}]
             \vtx{y}{-.9,0}           
           \end{scope}
           \arc{x}{u}
           \arc{u}{y}
           \vtx{z}{.9,0}
           \\
         };%
         \tikz[remember picture, overlay] \draw[->,thick](interface.north) --
         (up.south);%
         \tikz[remember picture, overlay] \draw[->,thick](interface.south) --
         (down.north);%
          }
        };
       \node[fill=white!79!black]
       (Y) at (0,-1)
       {%
         \parbox{.22\textwidth}{%
           \centering
           \tikz[remember picture]
           \node
           [matrix,thick,fill=white!99!black,column sep=1em,row sep=1em]
           (up)
           {
             \vtx{x}{0,0}            
              \begin{scope}[shift={(-.9,0)}]
                \phantom{\vtx{y}{-.9,0}}
              \end{scope}
              \vtx{z}{.9,0}
              \arc{x}{z}
             \\
           };%
           \\[1.1ex]
         \tikz[remember picture]
            \node
            [matrix,thick,fill=white!99!black,column sep=1em,row sep=1em]
            (interface)
            {
              \phantom{\vtx{u}{-.9,0}}
              \vtx{x}{0,0}
              \vtx{z}{.9,0}
              \phantom{%
                \begin{scope}[shift={(-.9,0)}]
                  \phantom{\vtx{y}{-.9,0}}
                \end{scope}
                \arc{x}{u}
                \arc{x}{z}
              }
              \\
            };
         \\[1.1ex]
         \tikz[remember picture]
         \node
         [matrix,thick,fill=white!99!black,column sep=1em,row sep=1em]
         (down)
         {
           \vtx{x}{0,0}
           \vtx{u}{-.9,0}
           \begin{scope}[shift={(-.9,0)}]
             \vtx{y}{-.9,0}           
           \end{scope}
           \arc{u}{y}
           \vtx{z}{.9,0}
           \\
         };%
         \tikz[remember picture, overlay] \draw[->,thick](interface.north) --
         (up.south);%
         \tikz[remember picture, overlay] \draw[->,thick](interface.south) --
         (down.north);%
          }
        };
        \node[fill=white!79!black]
       (Z) at (1,-1)
       {%
         \parbox{.22\textwidth}{%
           \centering
           \tikz[remember picture]
           \node
           [matrix,thick,fill=white!99!black,column sep=1em,row sep=1em]
           (up)
           {
             \vtx{v}{0,0}            
              \begin{scope}[shift={(-.9,0)}]
                \phantom{\vtx{y}{-.9,0}}
              \end{scope}
              \phantom{\vtx{z}{.9,0}}
              \begin{scope}[overlay]
                \arcC{v}{v}{.9,-.2}{.9,.2}{90}           
              \end{scope}
             \\
           };%
           \\[1.1ex]
         \tikz[remember picture]
            \node
            [matrix,thick,fill=white!99!black,column sep=1em,row sep=1em]
            (interface)
            {
              \phantom{\vtx{u}{-.9,0}}
              \vtx{v}{0,0}
              \phantom{\vtx{z}{.9,0}}
              \phantom{%
                \begin{scope}[shift={(-.9,0)}]
                  \phantom{\vtx{y}{-.9,0}}
                \end{scope}
              }
              \\
            };
         \\[1.1ex]
         \tikz[remember picture]
         \node
         [matrix,thick,fill=white!99!black,column sep=1em,row sep=1em]
         (down)
         {
           \vtx{v}{0,0}
           \vtx{u}{-.9,0}
           \begin{scope}[shift={(-.9,0)}]
             \vtx{y}{-.9,0}           
           \end{scope}
           \arc{u}{y}
           \phantom{\vtx{z}{.9,0}}
           \\
         };%
         \tikz[remember picture, overlay] \draw[->,thick](interface.north) --
         (up.south);%
         \tikz[remember picture, overlay] \draw[->,thick](interface.south) --
         (down.north);%
          }
        };
       \begin{scope}[shorten >=2pt,shorten <=2pt,]
         \draw [->] (Y) -- (Z);
         \draw [>->] (Y) -- (X);
       \end{scope}
       \foreach \u/\v in {L/X,K/Y,R/Z}
       \draw[shorten >=2pt,shorten <=2pt,>->] (\u) -- (\v);
     \end{tikzpicture}}
    \caption{Pushout decomposition of a transformation}
    \label{fig:PO-decomposition}
  \end{figure}
  To obtain a $({\cdot}\gets{\cdot}\to{\cdot})$-decomposition, 
  we chose $x$ and $z$ in the left hand side
  and $v$ -- the image of $x$ and $z$ -- in right hand side  as ``interfaces''.
  The two components of each object are very roughly  ``everything to the left'' of
  the ``interface'', 
  and ``everything to the right'' of the ``interface''.
  Due to the particular choice of the objects in this example,  
  the colimit construction, 
  which yields the objects of the original diagram, 
  is exactly the union of the components.
\end{example}

Our first theorem will be the soundness theorem for colimit
decompositions of arbitrary double pushout transformations. 
The involved notion of soundness of a double square approach~$\CAT{D}$
can be described roughly as follows:
whenever a double square diagram in the functor category
$[\CAT{J},\CAT{C}]$
consists of families of diagrams in $\CAT{D}$, 
we can combine the family of diagrams into one diagram in $\CAT{D}$ by taking colimits. 
The formal details are as follows.

\begin{definition}[Soundness]
  Let $\CAT{D}$ be a double square approach in $\CAT{C}$;
  it is  \emph{sound w.r.t.\ colimit decomposition}
  if for each category $\CAT{J}$ and each double square 
  diagram in~$[\CAT{J},\CAT{C}]$ as shown on the left below 
  \begin{displaymath}
          \begin{tikzpicture}[baseline={(bl)}]
      \mnode[L]{\mathcal L}{-1,1}
      \mnode[K]{\mathcal K}{0,1}
      \mnode[R]{\mathcal R}{1,1}
      \ardrawl{K}{L}{swap}{\alpha }
      \ardrawl{K}{R}{}{\beta}
      
      \mnode[X]{\mathcal X}{-1,0}
      \mnode[Y]{\mathcal Y}{0,0}
      \mnode[Z]{\mathcal Z}{1,0}
      \ardrawl{Y}{X}{}{\gamma}
      \ardrawl{Y}{Z}{swap}{\delta}
      \foreach \u/\v/\m/\s in {L/X/\mu/swap,K/Y/\iota/swap,R/Z/\nu/}
      \ardrawl{\u}{\v}{\s}{\m};
      \draw[draw=none] (R) -- (X) coordinate[pos=.5] (bl);
    \end{tikzpicture} 
    \qquad
      \begin{tikzpicture}[baseline={(bl)}]
      \mnode[L]{\mathcal L_i}{-1,1}
      \mnode[K]{\mathcal K_i}{0,1}
      \mnode[R]{\mathcal R_i}{1,1}
      \ardrawl{K}{L}{swap}{\alpha_i }
      \ardrawl{K}{R}{}{\beta_i}
      
      \mnode[X]{\mathcal X_i}{-1,0}
      \mnode[Y]{\mathcal Y_i}{0,0}
      \mnode[Z]{\mathcal Z_i}{1,0}
      \ardrawl{Y}{X}{}{\gamma_i}
      \ardrawl{Y}{Z}{swap}{\delta_i}
      \foreach \u/\v/\m/\s in {L/X/\mu_i/swap,K/Y/\iota_i/swap,R/Z/\nu_i/}
      \ardrawl{\u}{\v}{\s}{\m};
      \draw[draw=none] (R) -- (X) coordinate[pos=.5] (bl);
    \end{tikzpicture} 
  \end{displaymath}
  such that the right hand diagram in the above display belongs to
  $\CAT{D}$ 
  for each $i \in \CAT{J}$
  and colimits of $\mathcal{L,K,R,X,Y,}$ and $\mathcal{Z}$ exist, 
  the induced diagram, which is shown below,   also belongs to $\CAT{D}$. 
  \begin{displaymath}
    \begin{tikzpicture}[baseline={(bl)},scale=2,yscale=.5]
      \mnode[L]{\col \mathcal L}{-1,1}  
      \mnode[K]{\col \mathcal K}{0,1}
      \mnode[R]{\col \mathcal R}{1,1}
      \ardrawl{K}{L}{swap}{\col \alpha \vphantom{\beta}}
      \ardrawl{K}{R}{}{\col \beta}
      
      \mnode[X]{\col \mathcal X}{-1,0}
      \mnode[Y]{\col \mathcal Y}{0,0}
      \mnode[Z]{\col \mathcal Z}{1,0}
      \ardrawl{Y}{X}{}{\col \gamma}
      \ardrawl{Y}{Z}{swap}{\col \delta}
      \foreach \u/\v/\m/\s in {L/X/\mu/swap,K/Y/\iota/swap,R/Z/\nu/}
      \ardrawl{\u}{\v}{\s}{\col \m};
      \draw[draw=none] (R) -- (X) coordinate[pos=.5] (bl);
    \end{tikzpicture} 
  \end{displaymath}
\end{definition}
The next theorem
differs from the soundness theorem of~\cite{Rens10icalp}
in the following three ways:
first we use arbitrary double pushout rewriting;
second we use arbitrary colimits for decomposition;
finally, we do not impose restrictions on the natural transformations $\alpha$ and $\beta$ in
Definition~\ref{def:trans-decomp}. 
\begin{theorem}[Soundness]
  In categories that have (enough) pushouts, 
  double pushout rewriting is sound w.r.t.\ colimit decomposition.
\end{theorem}
\begin{proof}[Proof idea]
  If in the category of transformation, 
  the two spans   $(\col\mathcal{L}) \gets [\col
  \alpha] (\col \mathcal{K}) \to[\col \iota] (\col \mathcal{Y})$ 
  and $(\col \mathcal{Y}) \gets [\col \iota] (\col \mathcal{K}) \to
  [\col \beta] (\col \mathcal{R})$
  have pushouts, 
  then the desired follows   from the universal properties of  pushouts an colimits.
\end{proof}
\begin{remark}
  \label{rem:VKcolimitsRelation}
  We can obtain a variation of this soundness theorem 
  such that  (one of the) resulting rule morphisms 
  becomes a \mono. 
  (Such a result would be closer to the soundness theorem of~\cite{Rens10icalp}.) 
  For example, if the left morphism $\col \mathcal{L} \gets[\col
  \alpha] \col \mathcal{K}$
  should be a \mono, 
  we would make the following additional assumptions:
  first, 
  the colimit $\col \mathcal{L}$
  must be a Van Kampen colimit~\cite{Heindel2009},
  where Van Kampen colimits are the  generalization  of Van Kampen
  squares to colimits of arbitrary shape\footnote{%
    There is also the corresponding variation of
    Fact~\ref{fact:object-decomposition}
    that yields Van Kampen colimit decompositions of objects~\cite[Corollary~1]{BBHKlattice}.
  };
  second, 
  the natural transformation $\mathcal{L} \gets[\alpha] \mathcal{K}$ 
  would be required to be cartesian
  (see Definition~\ref{def:pb-stability-of-colimits}). 
\end{remark}

This theorem shows that there are enough double square diagrams that
are decomposable; 
the same proof idea can be used for  single pushout transformations of graphs
(and finite objects of locally cartesian closed adhesive categories).
However, 
the main topic of this paper concerns the ``converse'' of this
theorem and a more detailed analysis of situations
in which a double square transformation can be decomposed.

\section{The decomposition problem}
\label{sec:decomp-probl}
To motivate the decomposition problem for transformations, 
assume that we have a ``distributed representation'' of a system
state, namely an object $X$ that is the colimit $\col \mathcal{X}$ 
of some functor $\mathcal{X} \colon \CAT{J} \to \CAT{C}$
and thus for each $j \in \CAT{J}$ each ``component'' $\mathcal{X}_i$
could be stored on a different computer in a network.
Moreover, 
assume that we know  that there exists a transformation $X \tra[m]q
Y$, 
e.g.\ since we have an efficient (distributed) algorithm that determines that
$m$ is actually a match
(while avoiding the construction of the whole double square diagram).
Finally, 
we want to execute this transformation ``locally''
on each component $\mathcal{X}_i$ of $X$, 
e.g.\ because the construction of a ``global'' transformation
 $X \tra[m]q Y$ would be inefficient
if each component $\mathcal{X}_i$ could instead be processed independently
by one of the computers in the network.
In the end,
we want to find decompositions  of $m$ and $q$ 
such that the ``global'' transformation 
corresponds to a family of ``local'' transformations.
%
%
The formal description of the decomposition problem is as follows. 
\begin{definition}[Local decomposition problem]
  \label{def:a-priori}
  Let $\CAT{D}$ be a double square approach in $\CAT{C}$
  that is sound w.r.t.\ colimit decomposition;
  further let $\CAT{J}$ be a category. 

  A \emph{(local) $\CAT{J}$-decomposition problem} 
  is a triple $\langle \xi , q, m \rangle$
  where  $\xi \colon \mathcal{X} \cto X$ is a colimit of 
  a functor $\mathcal{X} \colon \CAT{J} \to \CAT{C}$, 
  $q = (L \gets[a] K \to[b] R)$ is a rule, 
  and  $m \colon L \to X$ is a match, 
  i.e.\ $\CAT{D}$ contains a diagram of the following form.
  \begin{equation}
    \label{eq:choice}
        \begin{tikzpicture}[baseline={(bl)}]
      \mnode{L}{-1,1}
      \mnode{K}{0,1}
      \mnode{R}{1,1}
      \ardrawl{K}{L}{swap}{a}
      \ardrawl{K}{R}{}{b}
      
      \mnode{X}{-1,0}
      \mnode{Y}{0,0}
      \mnode{Z}{1,0}
      \ardrawl{Y}{X}{}{c\vphantom{d}}
      \ardrawl{Y}{Z}{swap}{d}
      \foreach \u/\v/\m/\s in {L/X/m/swap,K/Y/j/swap,R/Z/n/}
      \ardrawl{\u}{\v}{\s}{\m};
      \draw[draw=none] (R) -- (X) coordinate[pos=.5] (bl);
    \end{tikzpicture}
  \end{equation}

  A \emph{solution} of a local $\CAT{J}$-decomposition problem
  $\langle \xi , q, m \rangle$
  is a pair $\langle \mu , \mathcal{Q}\rangle$
  where $\mathcal{Q} = \mathcal{L} \gets [\alpha] \mathcal{K}
  \to [\beta] \mathcal{R}$ is a rule in $[\CAT{J},\CAT{C}]$
  and $\mu \colon \mathcal{L} \to \mathcal{X}$ a natural transformation
  such that
  \begin{itemize}
  \item 
    the three equations $m = \col \mu$, $a = \col \alpha$, 
    and $b = \col \beta$ hold (for suitable choices of colimits of
    $\mathcal{L}$,$\mathcal{K}$, and $\mathcal{R}$);
  \item  $\mu$, $\alpha$, and $\beta$ can be 
    completed to the double square diagram 
    that is shown below on the left
    \begin{displaymath}
        \begin{tikzpicture}[baseline={(bl)}]
      \mnode[L]{\mathcal L}{-1,1}
      \mnode[K]{\mathcal K}{0,1}
      \mnode[R]{\mathcal R}{1,1}
      \ardrawl{K}{L}{swap}{\alpha }
      \ardrawl{K}{R}{}{\beta}
      
      \mnode[X]{\mathcal X}{-1,0}
      \mnode[Y]{\mathcal Y}{0,0}
      \mnode[Z]{\mathcal Z}{1,0}
      \ardrawl{Y}{X}{}{\gamma}
      \ardrawl{Y}{Z}{swap}{\delta}
      \foreach \u/\v/\m/\s in {L/X/\mu/swap,K/Y/\iota/swap,R/Z/\nu/}
      \ardrawl{\u}{\v}{\s}{\m};
      \draw[draw=none] (R) -- (X) coordinate[pos=.5] (bl);
    \end{tikzpicture} 
    \qquad \qquad
      \begin{tikzpicture}[baseline={(bl)}]
      \mnode[L]{\mathcal L_i}{-1,1}
      \mnode[K]{\mathcal K_i}{0,1}
      \mnode[R]{\mathcal R_i}{1,1}
      \ardrawl{K}{L}{swap}{\alpha_i }
      \ardrawl{K}{R}{}{\beta_i}
      
      \mnode[X]{\mathcal X_i}{-1,0}
      \mnode[Y]{\mathcal Y_i}{0,0}
      \mnode[Z]{\mathcal Z_i}{1,0}
      \ardrawl{Y}{X}{}{\gamma_i}
      \ardrawl{Y}{Z}{swap}{\delta_i}
      \foreach \u/\v/\m/\s in {L/X/\mu_i/swap,K/Y/\iota_i/swap,R/Z/\nu_i/}
      \ardrawl{\u}{\v}{\s}{\m};
      \draw[draw=none] (R) -- (X) coordinate[pos=.5] (bl);
    \end{tikzpicture}
  \end{displaymath}
  where  colimits of $\mathcal{Y}$ and  $\mathcal{Z}$ exist
  and moreover,
 for each $i\in \CAT{J}$, 
 the right one of the above diagrams belongs to $\CAT{D}$. 
  \end{itemize}
\end{definition}

Following the terminology of~\cite{Rens10icalp}, 
a rewriting approach is \emph{complete} w.r.t.\ to colimit decomposition
if each local decomposition problem has a (constructive) solution.
Moreover,  
the latter work gives a partial {completeness theorem}: 
in any adhesive category, 
local $({\cdot}\gets{\cdot}\to{\cdot})$-decomposition problems 
of a certain class have solutions in the double pushout approach. 

In this section, 
two similar partial completeness results will be given. 
The first one
applies to  any sound double square approach, 
and in particular to single-, double- and sesqui-pushout graph
transformation;
its proof is based on pullback stability of colimit decompositions
and the idea is to  ``lift''  decomposition of objects by means of
pullbacks. 
As it seems that the proof has rather the character of a brute force algorithm, 
we shall refine the basic proof idea
to devise a ``more local'' method
that provides solutions for a local decomposition problem with extra information. 
In fact, our second result will be a generalization of the completeness theorem
of~\cite{Rens10icalp}.

\pagebreak\subsection{Global decomposition}%
\nopagebreak%
\label{sec:global-decomposition}%
\nopagebreak%
\parpic[r]{\nopagebreak\noindent\nopagebreak
  \begin{tikzpicture}[baseline={(bl)}]
    \mnode{L}{-1,1}
    \mnode{K}{0,1}
    \mnode{R}{1,1}
    \ardrawl{K}{L}{swap}{a}
    \ardrawl{K}{R}{}{b}
    \mnode{X}{-1,0}
    \mnode{Y}{0,0}
    \mnode{Z}{1,0}
    \ardrawl{Y}{X}{}{c\vphantom{d}}
    \ardrawl{Y}{Z}{swap}{d}
    \foreach \u/\v/\m in {L/X/m,K/Y/,R/Z/}
    \ardrawl{\u}{\v}{swap}{\m};
    \draw[draw=none] (R) -- (X) coordinate[pos=.5] (bl);
  \end{tikzpicture}
}
We first present a procedure that can be used to split ``global''
reactions into families of ``local'' ones that act on the
constituents of states. 
The idea is to ``lift'' the colimit decompositions
of objects of Fact~\ref{fact:object-decomposition}  
to a double square transformation
by performing the following two constructions.
Given a double square diagram as shown to the right, 
we first assemble all involved entities of the reaction;
this can be achieved  by taking the pushout of $X \gets[c] Y \to[d]Z$, 
which yields an object $U$ that represents all entities that may be involved
in the reaction that is modelled by the transformation.
In a second step,
we take successive pullbacks to lift a decomposition of~$U$
to all other objects in the transformation diagram; 
in this way we obtain a family of double square diagrams, 
namely one for each irreducible component of the object~$U$.
This lifting procedure presupposes
that  the relevant double square approach is 
pullback stable. 
\begin{definition}[Pullback stability]
  \label{def:double-square-pb}
  Let $\CAT{C}$ be a category and let $\CAT{D}$ be a double square
  approach;
  further let $D \in \CAT{D}$ be a diagram
  as illustrated in Figure~\ref{fig:pb-lifting}. 
  \begin{figure}[htb]
    \centering
  \begin{tikzpicture}[scale=.9]
    \mnode{L}{-1,1}
    \mnode{K}{0,1}
    \mnode{R}{1,1}
    \ardrawl{K}{L}{swap}{a}
    \ardrawl{K}{R}{}{b}
    
    \mnode{X}{-1,0}
    \mnode{Y}{0,0}
    \mnode{Z}{1,0}
    \ardrawl{Y}{X}{}{}
    \ardrawl{Y}{Z}{swap}{}
    \foreach \u/\v/\m in {L/X/m,K/Y/,R/Z/}
    \ardrawl{\u}{\v}{}{\m};
    \begin{pgfonlayer}{background}
      \draw[draw=none,fill=lightgray] ($(L) + (-2.5ex,2.5ex)$) rectangle
      ($(Z) + (2.5ex,-2.5ex)$);
    \end{pgfonlayer}
    \draw[decorate,decoration={brace,mirror}]
    ($(L) + (-2.5ex,2.5ex)$)
    --
    node (D) [left]{$D\,$}
    ($(X) + (-2.5ex,-2.5ex)$); 
    \begin{scope}[shift={(0,2)}]
    \mnode[L]{L'}{-1,1}
    \mnode[K]{K'}{0,1}
    \mnode[R]{R'}{1,1}
    \ardrawl{K}{L}{swap}{a'}
    \ardrawl{K}{R}{}{b'}
    
    \mnode[X]{X'}{-1,0}
    \mnode[Y]{Y'}{0,0}
    \mnode[Z]{Z'}{1,0}
    \ardrawl{Y}{X}{}{}
    \ardrawl{Y}{Z}{swap}{}
    \foreach \u/\v/\m in {L/X/m',K/Y/,R/Z/}
    \ardrawl{\u}{\v}{}{\m};
    \begin{pgfonlayer}{background}
      \draw[draw=none,fill=lightgray] ($(L) + (-2.5ex,2.5ex)$) rectangle
      ($(Z) + (2.5ex,-2.5ex)$);
    \end{pgfonlayer}
    \draw[decorate,decoration={brace,mirror}]
    ($(L) + (-2.5ex,2.5ex)$)
    --
    node (D') [left]{$D'\,$}
    ($(X) + (-2.5ex,-2.5ex)$); 
    \end{scope}
    \ardrawl{D'}{D}{left}{\psi}
  \end{tikzpicture}\qquad\qquad
  \begin{tikzpicture}[scale=1.6]
    \mnode{L}{-1,0,0}
    \mnode{K}{0,0,0}
    \mnode{R}{1,0,0}
    \ardrawl{K}{L}{swap,pos=.8}{a}
    \ardrawl{K}{R}{pos=.2}{b}

    \mnode{X}{-1,0,1.25}
    \mnode{Y}{0,0,1.25}
    \mnode{Z}{1,0,1.25}
    \ardrawl{Y}{X}{}{}
    \ardrawl{Y}{Z}{swap}{}
    \foreach \u/\v/\m in {L/X/m,K/Y/,R/Z/}
    \ardrawl{\u}{\v}{}{\m};
    \begin{scope}[shift={(0,1.5,0)}]
      \mnode{L'}{-1,0,0}
    \mnode{K'}{0,0,0}
    \mnode{R'}{1,0,0}
    \mnode{X'}{-1,0,1.25}
    \mnode{Y'}{0,0,1.25}
    \mnode{Z'}{1,0,1.25}
    \POCS[.23]{K'}{-1,0,0}{0,-1.5,0}
    \POCS[.23]{K'}{1,0,0}{0,-1.5,0}
    \POCS[.23]{Y'}{-1,0,0}{0,-1.5,0}
    \POCS[.23]{Y'}{1,0,0}{0,-1.5,0}
    \POCS[.23]{R'}{0,-1.5,0}{0,0,1.25}
    \POCS[.23]{L'}{0,-1.5,0}{0,0,1.25}
    \POCSd[.23]{K'}{0,-1.5,0}{0,0,1.25}
    \foreach \v in {L,K,R,X,Y,Z}
    \ardrawld{\v'}{\v}{pos=.6}{\psi_{\v}};
    \ardrawld{Y'}{X'}{}{}
    \ardrawld{Y'}{Z'}{swap}{}
    \ardrawld{K'}{L'}{swap}{a'}
    \ardrawld{K'}{R'}{}{b'}
    \foreach \u/\v/\m in {L/X/m',K/Y/,R/Z/}
    \ardrawld{\u'}{\v'}{swap}{\m};
    \end{scope}
  \end{tikzpicture}
    \caption{Pullback lifting $D'$ of a double square diagram $D$}
    \label{fig:pb-lifting}
  \end{figure}
  A~\emph{pullback lifting of $D$}
  is a double square diagram $D'$ as shown in
  Figure~\ref{fig:pb-lifting}
  for which there exists a \emph{cartesian transformation}
  $\psi \colon D \to D'$,
  which means that there are morphisms $\psi_L, \psi_K, \psi_R, \psi_X,\psi_Y$,
  and $\psi_Z$ that yield pullback squares in the right hand
  diagram in Figure~\ref{fig:pb-lifting}.

  The approach $\CAT{D}$ is \emph{pullback stable}
  if any pullback lifting of any diagram in~$\CAT{D}$ is again a member of~$\CAT{D}$. 
\end{definition}
The above mentioned common approaches to graph transformation are all pullback stable
and we can apply the following proposition to decompose transformations.
\begin{proposition}[Global decomposition]
  \label{prop:global-decomposition}
  Let $\CAT{C}$ be a category with pullbacks,
  let $\CAT{D}$ be a pullback stable double square approach
  that is sound w.r.t.\ colimit decomposition,
  and let the left one of the 
  diagrams in~\eqref{eq:global} be a diagram in~$\CAT{D}$;
  moreover let $\CAT{J}$ be a category and $\mathcal{X} \colon \CAT{J}
  \to \CAT{C}$ be a functor with colimit $\xi \colon \mathcal{X} \cto
  X$. 
  \begin{equation}
    \label{eq:global}
    \begin{tikzpicture}[baseline={(bl)}]
      \mnode{L}{-1,1}
      \mnode{K}{0,1}
      \mnode{R}{1,1}
      \ardrawl{K}{L}{swap}{a}
      \ardrawl{K}{R}{}{b}
      
      \mnode{X}{-1,0}
      \mnode{Y}{0,0}
      \mnode{Z}{1,0}
      \ardrawl{Y}{X}{}{c\vphantom{d}}
      \ardrawl{Y}{Z}{swap}{d}
      \foreach \u/\v/\m/\s in {L/X/m/swap,K/Y/j/swap,R/Z/n/}
      \ardrawl{\u}{\v}{\s}{\m};
      \draw[draw=none] (R) -- (X) coordinate[pos=.5] (bl);
    \end{tikzpicture}
    \qquad
    \begin{tikzpicture}[baseline={(bl)}]
      \mnode{L}{-1,1}
      \mnode{K}{0,1}
      \mnode{R}{1,1}
      \ardrawl{K}{L}{swap}{a}
      \ardrawl{K}{R}{}{b}
      
      \mnode{X}{-1,0}
      \mnode{Y}{0,0}
      \mnode{Z}{1,0}
      \ardrawl{Y}{X}{}{c\vphantom{d}}
      \ardrawl{Y}{Z}{swap}{d}
      \foreach \u/\v/\m/\s in {L/X/m/swap,K/Y/j/swap,R/Z/n/}
      \ardrawl{\u}{\v}{\s}{\m};
      \draw[draw=none] (R) -- (X) coordinate[pos=.5] (bl);
      \mnode{U}{-1,-1}
      \ardrawl{X}{U}{swap}{w}
      \ardrawl[{to[bend left=30]}]{Z}{U}{}{t}
      \begin{scope}[shift={(.5,0)}]
        \mnode[X']{\mathcal X_i}{-3,0}
        \ardrawl{X'}{X}{}{\xi_i}
        \mnode[U']{\mathcal{U}_i}{-3,-1}
        \ardrawl{U'}{U}{swap}{\varphi_i}
        \ardrawl{X'}{U'}{swap}{\omega_i}
        \POCS[.3]{X'}{1.5,0}{0,-1}
      \end{scope}
    \end{tikzpicture}
  \end{equation}

  The local decomposition problem 
  $\bigl\langle \xi,\, (L \gets[a] K \to[b] R) ,\,  m\bigr\rangle$
  has a solution if there exists a cospan $X \to[w] U \gets [t] Z$
  that satisfies $w \circ c = t \circ d$
  and moreover there exists a diagram $\mathcal{U} \colon \CAT{J} \to \CAT{C}$ 
  with a pullback stable colimit $\varphi \colon \mathcal{U} \cto U$ 
  such that $\xi$ is the pullback of $\varphi$ along $w$, 
  i.e.\ for each $i \in \CAT{J}$, 
  there exists a morphism
  $\omega_i \colon \mathcal{X}_i \to  \mathcal{U}_i$
  that makes 
  $\mathcal{U}_i \gets [\omega_i] \mathcal{X}_i \to[\xi_i] X$  
  a pullback of $\mathcal{U}_i \to[\varphi_i] U \gets[w] X$ 
  (as illustrated in~\eqref{eq:global} on the right).
\end{proposition}

\begin{proof}[Proof sketch]
  Since the colimit $\mathcal{U}$ is pullback stable
  and the right hand diagram in~\eqref{eq:global}
  commutes, 
  we can form successive pullbacks to obtain 
  a colimit decomposition
  of the left hand double square diagram 
  in~\eqref{eq:global}. 
  The desired then follows from the assumption that 
  $\CAT{D}$ is pullback stable. 
\end{proof}

It is obvious, 
that (the proof of) this proposition is not of much use 
if we want to solve the local decomposition problem efficiently, 
e.g.\ in the category of graphs. 
Even if we construct the cospan $X \to[w] U \gets[t] Z$ as a pushout, 
there are in general too many choices for suitable colimit decompositions
of~$U$.
Also, this approach is fully ``global'' as we need to construct the
whole double square diagram in $\CAT{C}$
and do not only manipulate its components ``locally''.

Moreover, 
note that the proof of Proposition~\ref{prop:global-decomposition}
only makes use of colimit decompositions of a particular type:
all transformation decompositions in the functor category $[\CAT{J},\CAT{C}]$
consists of cartesian transformations.
Hence, 
if we consider only  colimit decompositions of this kind,
there might be a more elegant method for the solution of local
decomposition problems.
In fact, with some extra information on a local decomposition problem, 
we can devise a purely local decomposition method.

\subsection{More local decompositions}
\label{sec:constr-decomp}
We have seen that some decomposition problems can be solved
by employing the notion of pullback lifting of double square diagrams.
We  shall now  make use of the good behaviour of pushouts w.r.t.\ to
pullbacks in  adhesive categories
to develop a ``local'' method for the solution
of a simplified decomposition problem for double pushout transformations. 
Although the full details are somewhat technical, 
the central point is the following lemma, 
which follows directly from the definition of adhesive categories. 

\begin{lemma}[Double Pushout Lifting]
  \label{lem:local-lifting}
  Let $\CAT{C}$ be an adhesive category.
  Given a diagram as shown in Figure~\ref{fig:dpo-lifting} on the left
  \begin{figure}[htb]
    \centering
      \begin{tikzpicture}[scale=1.5,yscale=.9]
    \mnode{L}{-1,0,0}
    \mnode{K}{0,0,0}
    \mnode{R}{1,0,0}
    \ardrawl{K}{L}{swap,pos=.8}{a}
    \ardrawl{K}{R}{pos=.2}{b}

    \mnode{X}{-1,0,1.25}
    \mnode{Y}{0,0,1.25}
    \mnode{Z}{1,0,1.25}
    \ardrawl{Y}{X}{}{}
    \ardrawl{Y}{Z}{swap}{}
    \foreach \u/\v/\m in {L/X/m,K/Y/,R/Z/}
    \ardrawl{\u}{\v}{swap,pos=.3}{\m};
    \begin{scope}[shift={(0,1.5,0)}]
      \mnode{L'}{-1,0,0}
    \mnode{K'}{0,0,0}
    \mnode{R'}{1,0,0}
    \mnode{X'}{-1,0,1.25}
    \POCS[.23]{K'}{-1,0,0}{0,-1.5,0}
    \POCS[.23]{K'}{1,0,0}{0,-1.5,0}
    \POCS[.23]{L'}{0,-1.5,0}{0,0,1.25}
    \foreach \v in {L,K,R,X
    }
    \ardrawld{\v'}{\v}{pos=.9,swap}{\psi_{\v}};
    \ardrawld{K'}{L'}{swap}{a'}
    \ardrawld{K'}{R'}{}{b'}
    \foreach \u/\v/\m in {L/X/m'
    }
    \ardrawld{\u'}{\v'}{swap}{\m};
    \POCS[.4]{X}{1,0}{0,0,-1.25}
    \POCS[.4]{Z}{-1,0}{0,0,-1.25}
    \end{scope}
  \end{tikzpicture}
  \qquad
  \begin{tikzpicture}[scale=1.5,yscale=.9]
    \mnode{L}{-1,0,0}
    \mnode{K}{0,0,0}
    \mnode{R}{1,0,0}
    \ardrawl{K}{L}{swap,pos=.8}{a}
    \ardrawl{K}{R}{pos=.2}{b}

    \mnode{X}{-1,0,1.25}
    \mnode{Y}{0,0,1.25}
    \mnode{Z}{1,0,1.25}
    \ardrawl{Y}{X}{}{}
    \ardrawl{Y}{Z}{swap}{}
    \foreach \u/\v/\m in {L/X/m,K/Y/,R/Z/}
    \ardrawl{\u}{\v}{swap,pos=.3}{\m};
    \begin{scope}[shift={(0,1.5,0)}]
      \mnode{L'}{-1,0,0}
    \mnode{K'}{0,0,0}
    \mnode{R'}{1,0,0}
    \mnode{X'}{-1,0,1.25}
    \mnode{Y'}{0,0,1.25}
    \mnode{Z'}{1,0,1.25}
     \POCS[.23]{Y'}{-1,0,0}{0,-1.5,0}
     \POCS[.23]{Y'}{1,0,0}{0,-1.5,0}
    \POCS[.23]{R'}{0,-1.5,0}{0,0,1.25}
    \POCSd[.23]{K'}{0,-1.5,0}{0,0,1.25}
    \foreach \v in {L,K,R,X,Y,Z}
    \ardrawld{\v'}{\v}{pos=.9,swap}{\psi_{\v}};
     \ardrawld{Y'}{X'}{}{}
     \ardrawld{Y'}{Z'}{swap}{}
    \ardrawld{K'}{L'}{swap}{a'}
    \ardrawld{K'}{R'}{}{b'}
    \foreach \u/\v/\m in {L/X/m',K/Y/,R/Z/}
    \ardrawld{\u'}{\v'}{swap}{\m};
    \POCSd[.4]{X'}{1,0}{0,0,-1.25}
    \POCSd[.4]{Z'}{-1,0}{0,0,-1.25}
    \end{scope}
  \end{tikzpicture}
    \caption{Double Pushout Lifting}
    \label{fig:dpo-lifting}
  \end{figure}
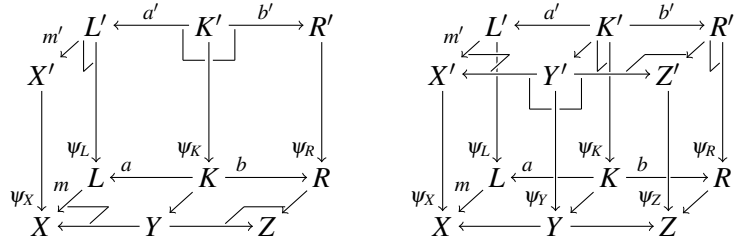
  where $a$ and $m$ or $a$ and $b$ are \mono, 
  the bottom squares are pushouts
  and the vertical faces are pullbacks, 
  there exists an extension of this diagram 
  as shown on the right in Figure~\ref{fig:dpo-lifting}, 
  which has two pushout squares as top faces 
  and all new lateral faces are pullbacks. 
\end{lemma}
\begin{proof}[Proof sketch]
  The left hand ``cube'' can be constructed by taking pullbacks
  and thus the left square on top must be a pushout
  by the defining property of adhesive categories. 
  The second pushout is obtained by taking a pushout 
  of $b'$ and the new morphism with domain $K'$. 
  Using the definition  of adhesive categories once more, 
  we derive that the two remaining new faces are pullbacks as well. 
\end{proof}
Note that this lemma is really directly related to the definition of adhesive
categories.
\
It is the main tool in  the proof of the completeness
theorem that we shall present below
and it is also closely related to the accommodations
that are used in~\cite{Rens10icalp} to obtain 
completeness of double pushout rewriting (w.r.t.\ pushout decompositions). 
Following  the terminology of the latter work, 
we define a class of local decomposition problems
that can be solved by means of Double Pushout Lifting. 
\begin{definition}[Accommodations for decomposition problems]
  \label{def:accomodations}
  Let
  $\langle \xi , (L \gets[a] K \to[b]  R), m  \rangle$
  be a local $\CAT{J}$-decomposition problem where $\xi \colon
  \mathcal{X} \cto X$ and $m \colon L \to X$.

  An \emph{accommodation (scheme) for $\langle \xi , (L \gets[a] K \to[b]  R), m  \rangle$}
  is a  pullback stable colimit
  $\rho \colon \mathcal{R} \cto  R$
  such that pulling back $\rho$ along $b$ 
  and pulling back $\xi$ along $m \circ a$ yield
  isomorphic results, i.e.\
  for each $i \in \CAT{J}$, 
  there is a diagram  of the following form.
  \begin{equation}\label{eq:accomodation}
    \begin{tikzpicture}[scale=1.7,yscale=.6]
    \mnode{L}{-1,0,0}
    \mnode{K}{0,0,0}
    \mnode{R}{1,0,0}
    \ardrawl{K}{L}{swap,pos=.8}{a}
    \ardrawl{K}{R}{pos=.2}{b}

    \mnode{X}{-1,0,1.25}
    \foreach \u/\v/\m in {L/X/m
    }
    \ardrawl{\u}{\v}{swap,pos=.3}{\m};
    \begin{scope}[shift={(0,1.5,0)}]
    \mnode[K']{\mathcal{K}_i'}{0,0,0}
    \mnode[R']{\mathcal{R}_i}{1,0,0}
    \mnode[X']{\mathcal{X}_i}{-1,0,1.25}
    \POCS[.23]{K'}{1,0,0}{0,-1.5,0}
    \foreach \v/\m in {
      K/,R/\rho_i,X/\xi_i
    }
    \ardrawld{\v'}{\v}{pos=.2}{\m};
    \ardrawld{K'}{R'}{}{}
    \ardrawl{K}{X}{}{m \circ a}
    \ardrawl{K'}{X'}{}{}
    \POCS[.23]{K'}{0,-1.5,0}{-1,0,1.25}
    \end{scope}
  \end{tikzpicture}
  \end{equation}
\end{definition}

In~\cite{Rens10icalp},
attention is restricted to local $\CAT{J}$-decomposition problems
where $\CAT{J}$ is the shape of spans,
i.e.\  $\CAT{J}$ is fixed to be the category $({\cdot} \gets {\cdot}
\to {\cdot})$;
in fact, the object $X$ in Definition~\ref{def:accomodations} is
assumed to arise as the pushout of a pair of \monos and thus is a Van Kampen square. 
A reader that is familiar with the latter work, 
will notice that accommodation schemes correspond to suitable pairs of accommodations in the
sense of~\cite[Definition~9]{Rens10icalp}
and will be able to verify that the following theorem is a
generalization of the Completeness Theorem of~\cite{Rens10icalp};
the details are omitted due to space constraints (see also Remark~\ref{rem:VKcolimitsRelation}). 

\begin{theorem}[Accommodated Completeness]
  \label{thm:completeness}
  Let $\CAT{C}$ be an adhesive category, 
  let $\langle \xi , (L \gets[a] K \to[b]  R), m  \rangle$
  be a local  $\CAT{J}$-decomposition problem for the $\CAT{DPO}$ approach
  where at least one of  $m$ and $b$ is a \mono;
  then $\langle \xi , (L \gets[a] K \to[b]  R), m  \rangle$ has as solution.
\end{theorem}
\begin{proof}[Proof sketch]
  First we construct successive pullbacks of $\xi$ along $m$ and $a$, 
  which yields cartesian morphisms $\mu \colon \mathcal{L} \to
  \mathcal{X}$
  and $\alpha \colon \mathcal{K} \to \mathcal{L}$, respectively. 
  Since $\rho$ is an accommodation, 
  there is a cartesian transformation $\beta \colon \mathcal{K} \to \mathcal{R}$. 
 The candidate for a solution is $\langle \mu , \mathcal{L}
  \gets[\alpha] \mathcal{K}\to[\beta]\mathcal{R}\rangle$. 
  Finally, we apply 
  the Double Pushout Lifting Lemma and soundness of $\CAT{DPO}$
  to complete the proof.
\end{proof}

If we assume that in the situation of Diagram~\eqref{eq:accomodation} in Definition~\ref{def:accomodations}, 
the right hand side~$R$ is relatively small in comparison with the
state~$X$, 
the decomposition procedure that is implicitly given by the proof of Theorem~\ref{thm:completeness}
is more local than the global one of Section~\ref{sec:global-decomposition}
insofar as we ``only'' need to find a suitable accommodation of
the right hand side $R$ of the rule $L \gets[a] K \to[b] R$. 
More concretely, if we work in the category of graphs and use the
canonical decompositions of Section~\ref{sec:decomp-objects}, 
the object $\mathcal{K}_i'$ 
will be the empty graph for many $i \in \CAT{J}$
and then the typical first candidate for $\mathcal{R}_i$  is also the
empty graph;
it remains to be explored how good this heuristics is in practice. 

\section{Conclusion}
\label{sec:conclusion}
We have shown how arbitrary 
colimits can be used to decompose transformations of graph-like
objects
and 
 have generalized the results of~\cite{Rens10icalp}
in this setting:
first, we have given a soundness theorem for double pushout rewriting
that allows to combine families of ``local'' transformations into a
single ``global'' one;
moreover, 
our completeness theorem provides solutions for
\emph{accommodated} decomposition problems in adhesive categories;
finally, we have given a precise formulation of a general local decomposition
problem that can serve as a  base for future research
for double square transformation approaches 
that are used to model reactions of graph-like objects. 

The first,
immediate direction for future work is the description of the 
soundness (and completeness) theorem in a more standard \textsc{sos}
style using (an extension of) the de Simone format 
with terms over a suitable signature.
If we take the ``closed system'' perspective of the present paper
and further restrict to (de-)composition by means of (binary)
coproducts then the solution seems rather straightforward: 
if we have two ``local'' transitions  $X_1 \tra[m_1]{q_1} Y_i$ and
$X_2 \tra[m_2]{q_2} Y_2$
then $X_1 + X_2 \tra[m_1 + m_2]{q_1 + q_2} Y_1 + Y_2$ 
is the corresponding global transformation; 
however, even in the simple case of coproducts it is not quite clear
how the fact that the ``constants'' of the signature are actually
interpreted as (irreducible) objects of some adhesive category features. 

Besides the problem of a ``classical'' \textsc{sos} account of the
results of this paper, 
we are confronted with further issues. 
For example, it is very desirable to make the step from the closed
systems perspective of double pushout rewriting
to borrowed context rewriting~\cite{EH06,BEK06}, which captures essential
aspects of the interaction of open systems with their environment
in the form of labelled transition systems
that are automatically derived from rules of reaction.

Moreover, 
we want to reconsider the problem of reaction \emph{rates}
(which is already non-trivial in closed systems). 
The relevant problem in stochastic graph transformation
\cite{DBLP:journals/fuin/HeckelLM06}
and the $\kappa$-calculus\cite{DBLP:journals/tcs/DanosL04}
are the numerous symmetries of objects that have to be considered 
to obtain adequate quantitative system models.
To  ``deconstruct'' these symmetries, 
we plan to use the canonical colimit decomposition of finite objects in
adhesive categories. 
The goal is a distributed method 
for the computation of all isomorphisms of a given object
and we also hope to gain fundamental insights into the ``internal'' structure
of finite objects in adhesive categories.

\paragraph{Acknowledgements}
I would like to thank Arend Rensink for discussions about the main
ideas of his work on decomposition of graph transformations
and Barbara K{\"o}nig for inspiring comments on a draft version of
this paper;
moreover I would like to express my gratitude for the very constructive and
benevolent comments of the referees of this paper.

\bibliographystyle{eptcs}
\bibliography{sos2010}

\begin{thebibliography}{10}
\providecommand{\bibitemstart}[1]{\bibitem{#1}}
\providecommand{\bibitemend}{}
\providecommand{\bibliographystart}{}
\providecommand{\bibliographyend}{}
\providecommand{\url}[1]{\texttt{#1}}
\providecommand{\urlprefix}{Available at }
\providecommand{\bibinfo}[2]{#2}
\bibliographystart

\bibitemstart{BBHKlattice}
\bibinfo{author}{Paolo Baldan}, \bibinfo{author}{Filippo Bonchi},
  \bibinfo{author}{Andrea Corradini}, \bibinfo{author}{Tobias Heindel} \&
  \bibinfo{author}{Barbara K{\"o}nig} (\bibinfo{year}{To appear}):
  \emph{\bibinfo{title}{{A Lattice-Theoretical Perspective on Adhesive
  Categories}}}.
\newblock {\sl \bibinfo{journal}{Journal of Symbolic Computation}} .
\bibitemend

\bibitemstart{BEK06}
\bibinfo{author}{Paolo Baldan}, \bibinfo{author}{Hartmut Ehrig} \&
  \bibinfo{author}{Barbara K\"{o}nig} (\bibinfo{year}{2006}):
  \emph{\bibinfo{title}{Composition and Decomposition of {DPO} Transformations
  with Borrowed Context}}.
\newblock In: {\sl \bibinfo{booktitle}{Proc. of ICGT '06 (International
  Conference on Graph Transformation)}}, \bibinfo{publisher}{Springer}, pp.
  \bibinfo{pages}{153--167}.
\newblock \bibinfo{note}{LNCS 4178}.
\bibitemend

\bibitemstart{boehm1987amalgamation}
\bibinfo{author}{Paul Boehm}, \bibinfo{author}{Harald-Reto Fonio} \&
  \bibinfo{author}{Annegret Habel} (\bibinfo{year}{1987}):
  \emph{\bibinfo{title}{{Amalgamation of graph transformations: a
  synchronization mechanism}}}.
\newblock {\sl \bibinfo{journal}{Journal of Computer and System Sciences}}
  \bibinfo{volume}{34}(\bibinfo{number}{2-3}), pp. \bibinfo{pages}{377--408}.
\bibitemend

\bibitemstart{DBLP:confggCorradiniHHK06}
\bibinfo{author}{Andrea Corradini}, \bibinfo{author}{Tobias Heindel},
  \bibinfo{author}{Frank Hermann} \& \bibinfo{author}{Barbara K{\"o}nig}
  (\bibinfo{year}{2006}): \emph{\bibinfo{title}{{Sesqui-Pushout Rewriting}}}.
\newblock In: {\sl \bibinfo{booktitle}{Graph Transformations, Third
  International Conference, {ICGT} 2006, Natal, Rio Grande do Norte, Brazil,
  September 17-23, 2006, Proceedings}}, pp. \bibinfo{pages}{30--45}.
\bibitemend

\bibitemstart{DBLP:conf/gg/CorradiniMREHL97}
\bibinfo{author}{Andrea Corradini}, \bibinfo{author}{Ugo Montanari},
  \bibinfo{author}{Francesca Rossi}, \bibinfo{author}{Hartmut Ehrig},
  \bibinfo{author}{Reiko Heckel} \& \bibinfo{author}{Michael L{\"o}we}
  (\bibinfo{year}{1997}): \emph{\bibinfo{title}{{Algebraic Approaches to Graph
  Transformation -- Part {\sc I}: Basic Concepts and Double Pushout
  Approach}}}.
\newblock In \bibinfo{editor}{Rozenberg}  \cite{DBLP:conf/gg/1997handbook}, pp.
  \bibinfo{pages}{163--246}.
\bibitemend

\bibitemstart{DBLP:journals/tcs/DanosL04}
\bibinfo{author}{Vincent Danos} \& \bibinfo{author}{Cosimo Laneve}
  (\bibinfo{year}{2004}): \emph{\bibinfo{title}{{F}ormal {M}olecular
  {B}iology}}.
\newblock {\sl \bibinfo{journal}{Theoretical Computer Science}}
  \bibinfo{volume}{325}(\bibinfo{number}{1}), pp. \bibinfo{pages}{69--110}.
\bibitemend

\bibitemstart{EH06}
\bibinfo{author}{Hartmut Ehrig} \& \bibinfo{author}{Barbara K\"onig}
  (\bibinfo{year}{2006}): \emph{\bibinfo{title}{{Deriving Bisimulation
  Congruences in the {DPO} Approach to Graph Rewriting with Borrowed
  Contexts}}}.
\newblock {\sl \bibinfo{journal}{Mathematical Structures in Computer Science}}
  \bibinfo{volume}{16}(\bibinfo{number}{6}), pp. \bibinfo{pages}{1133--1163}.
\bibitemend

\bibitemstart{DBLP:conf/focs/EhrigPS73}
\bibinfo{author}{Hartmut Ehrig}, \bibinfo{author}{Michael Pfender} \&
  \bibinfo{author}{Hans~J{\"u}rgen Schneider} (\bibinfo{year}{1973}):
  \emph{\bibinfo{title}{{Graph-Grammars: An Algebraic Approach}}}.
\newblock In \bibinfo{editor}{IEEE}  \cite{DBLP:conf/focs/FOCS14}, pp.
  \bibinfo{pages}{167--180}.
\bibitemend

\bibitemstart{goguen1973categorical}
\bibinfo{author}{Joseph Goguen} (\bibinfo{year}{1973}):
  \emph{\bibinfo{title}{{Categorical foundations for general systems theory}}}.
\newblock {\sl \bibinfo{journal}{Advances in Cybernetics and Systems Research}}
  \bibinfo{volume}{1}.
\bibitemend

\bibitemstart{DBLP:journals/fuin/HeckelLM06}
\bibinfo{author}{Reiko Heckel}, \bibinfo{author}{Georgios Lajios} \&
  \bibinfo{author}{Sebastian Menge} (\bibinfo{year}{2006}):
  \emph{\bibinfo{title}{{Stochastic Graph Transformation Systems}}}.
\newblock {\sl \bibinfo{journal}{Fundamenta Informaticae}}
  \bibinfo{volume}{74}(\bibinfo{number}{1}), pp. \bibinfo{pages}{63--84}.
\bibitemend

\bibitemstart{Heindel2009}
\bibinfo{author}{Tobias Heindel} \& \bibinfo{author}{Pawe{\l} Soboci\'{n}ski}
  (\bibinfo{year}{2009}): \emph{\bibinfo{title}{{Van Kampen colimits as
  bicolimits in Span}}}.
\newblock In: {\sl \bibinfo{booktitle}{Algebra and Coalgebra in Computer
  Science: Third International Conference, Calco 2009, Udine, Italy, September
  7--10, 2009, Proceedings}}, number \bibinfo{number}{5728} in
  \bibinfo{series}{LNCS}, \bibinfo{publisher}{Springer}, pp.
  \bibinfo{pages}{335--349}.
\bibitemend

\bibitemstart{DBLP:conf/focs/FOCS14}
\bibinfo{editor}{IEEE}, editor (\bibinfo{year}{1973}):
  \emph{\bibinfo{title}{14th Annual Symposium on Foundations of Computer
  Science, 15-17 October 1973, The University of Iowa, USA}}.
  \bibinfo{publisher}{IEEE Computer Society Press}.
\bibitemend

\bibitemstart{lack:adhes}
\bibinfo{author}{Stephen Lack} \& \bibinfo{author}{Pawe{\l} Soboci{\'n}ski}
  (\bibinfo{year}{2005}): \emph{\bibinfo{title}{{Adhesive and Quasiadhesive
  Categories}}}.
\newblock {\sl \bibinfo{journal}{Theoretical Informatics and Applications}}
  \bibinfo{volume}{39}(\bibinfo{number}{2}), pp. \bibinfo{pages}{511--546}.
\bibitemend

\bibitemstart{solos}
\bibinfo{author}{Cosimo Laneve}, \bibinfo{author}{Joachim Parrow} \&
  \bibinfo{author}{Bj{\"o}rn Victor} (\bibinfo{year}{2001}):
  \emph{\bibinfo{title}{{Solo Diagrams}}}.
\newblock In: {\sl \bibinfo{booktitle}{Proceedings of the 4th International
  Symposium on Theoretical Aspects of Computer Software}},
  \bibinfo{organization}{Springer-Verlag}, pp. \bibinfo{pages}{127--144}.
\bibitemend

\bibitemstart{DBLP:journals/tcs/Lowe93}
\bibinfo{author}{Michael L{\"o}we} (\bibinfo{year}{1993}):
  \emph{\bibinfo{title}{{Algebraic Approach to Single-Pushout Graph
  Transformation}}}.
\newblock {\sl \bibinfo{journal}{Theoretical Computer Science}}
  \bibinfo{volume}{109}(\bibinfo{number}{1{\&}2}), pp.
  \bibinfo{pages}{181--224}.
\bibitemend

\bibitemstart{MacLaneS:catwm}
\bibinfo{author}{Saunders Mac~Lane} (\bibinfo{year}{1971}):
  \emph{\bibinfo{title}{{Categories for the Working Mathematician}}}.
\newblock Number~\bibinfo{number}{5} in \bibinfo{series}{Graduate Texts in
  Mathematics}. \bibinfo{publisher}{Springer}.
\bibitemend

\bibitemstart{PIERCE91}
\bibinfo{author}{Benjamin~C. Pierce} (\bibinfo{year}{1991}):
  \emph{\bibinfo{title}{{Basic Category Theory for Computer Scientists}}}.
\newblock \bibinfo{publisher}{MIT Press}.
\bibitemend

\bibitemstart{Rens10icalp}
\bibinfo{author}{Arend Rensink} (\bibinfo{year}{2010}):
  \emph{\bibinfo{title}{{Compositionality in Graph Transformation}}}.
\newblock In: \bibinfo{editor}{Samson Abramsky}, \bibinfo{editor}{Cyril
  Gavoille}, \bibinfo{editor}{Claude Kirchner},
  \bibinfo{editor}{Friedhelm~Meyer auf~der Heide} \& \bibinfo{editor}{Paul~G.
  Spirakis}, editors: {\sl \bibinfo{booktitle}{ICALP (2)}}, {\sl
  \bibinfo{series}{Lecture Notes in Computer Science}} \bibinfo{volume}{6199},
  \bibinfo{publisher}{Springer}, pp. \bibinfo{pages}{309--320}.
\newblock \urlprefix\url{http://dx.doi.org/10.1007/978-3-642-14162-1_26}.
\bibitemend

\bibitemstart{DBLP:conf/gg/1997handbook}
\bibinfo{editor}{Grzegorz Rozenberg}, editor (\bibinfo{year}{1997}):
  \emph{\bibinfo{title}{{Handbook of Graph Grammars and Computing by Graph
  Transformations, Volume 1: Foundations}}}. \bibinfo{publisher}{World
  Scientific}.
\bibitemend

\bibitemstart{DBLP:journals/njc/SassoneS03}
\bibinfo{author}{Vladimiro Sassone} \& \bibinfo{author}{Pawe{\l}
  Soboci{\'n}ski} (\bibinfo{year}{2003}): \emph{\bibinfo{title}{{Deriving
  Bisimulation Congruences using 2-categories}}}.
\newblock {\sl \bibinfo{journal}{Nordic Journal of Computing}}
  \bibinfo{volume}{10}(\bibinfo{number}{2}), pp. \bibinfo{pages}{163--183}.
\bibitemend

\bibitemstart{GabiPhD}
\bibinfo{author}{Gabriele Taentzer} (\bibinfo{year}{1996}):
  \emph{\bibinfo{title}{{Parallel and distributed graph transformation: Formal
  description and application to communication-based systems}}}.
\newblock \bibinfo{type}{Ph.D. thesis}, \bibinfo{school}{Technische
  Universit{\"a}t Berlin}.
\bibitemend

\bibitemstart{xi-calc}
\bibinfo{author}{Lucian Wischik} \& \bibinfo{author}{Philippa Gardner}
  (\bibinfo{year}{2005}): \emph{\bibinfo{title}{{Explicit fusions}}}.
\newblock {\sl \bibinfo{journal}{Theoretical Computer Science}}
  \bibinfo{volume}{340}(\bibinfo{number}{3}), pp. \bibinfo{pages}{606--630}.
\bibitemend

\bibliographyend
\end{thebibliography}
\end{document}